\documentclass{aa}
\makeatletter
\renewcommand*\aa@pageof{, page \thepage{} of \pageref*{LastPage}} 
\usepackage[varg]{txfonts}
\usepackage{natbib}
\bibpunct{(}{)}{;}{a}{}{,} 
\usepackage{hyperref}
\usepackage{enumitem}
\usepackage{subcaption}
\usepackage{graphicx}
\usepackage{siunitx} 
\usepackage{xcolor}
\usepackage{amsmath} 
\newcommand{\mgg}[1]{{#1}}
\newcommand{\mmgg}[1]{{#1}}
\newcommand{\mmg}[1]{{#1}}
\newcommand{\mg}[1]{{#1}}
\newcommand{\gm}[1]{{#1}}
\newcommand{\gmnew}[1]{{#1}}

\newcommand{\rf}[1]{#1}
\newcommand{\rff}[1]{#1}
\newcommand{\rrff}[1]{#1}
\newcommand{\nrf}[1]{#1}

\newcommand{\mrf}[1]{{#1}}
\newcommand{\mb}[1]{#1}
\newcommand{\mbf}[1]{#1}
\newcommand{\nbf}[1]{#1}

\usepackage{makecell} 

\DeclareMathAlphabet{\mathitbf}{OML}{cmm}{b}{it}

\newcommand{\Bz}{\mathit{B_z}}

\newcommand{\hpj}{H_{\mathrm{PJ}}}
\newcommand{\hj}{H_{\mathrm{J}}}
\newcommand{\hjprime}{|\hj|/|\hv|}
\newcommand{\hv}{H_{V}}
\newcommand{\mhjprime}{\langle|\hj|/|\hv|\rangle}

\newcommand{\degree}{^\circ}

\newcommand{\nc}{n_{\mathrm{crit}}}

\newcommand{\nmedian}{\widetilde{n}}
\newcommand{\nbar}{\overline n}
\newcommand{\hc}{h_{\mathrm{crit}}}
\newcommand{\hcmedian}{{\widetilde h_{\mathrm{crit}}}}
\newcommand{\hcbar}{{\overline h_{\mathrm{crit}}}}
\newcommand{\hcang}{{\langle\overline h_{\mathrm{crit}}\rangle}}

\newcommand{\hcenter}{h_{\mathrm{center}}}

\newcommand{\hcentermean}{{\overline h_{\mathrm{center}}}}
\newcommand{\deltah}{\Delta_{h}}

\newcommand{\mgplots}[7]{
    \hspace{6em}\vspace{0.5em}
    \begin{subfigure}[ht]{0.33\textwidth}
        \centering
        \begin{picture}(100,100)
        \put(0,0){\includegraphics[width=1.0\linewidth]{plots_new//#4//noaa_#1_#3#5}}
        {\put(32,96){#6) AR #1}}
        \end{picture}
    \end{subfigure}\hspace{1em}
    \begin{subfigure}[ht]{0.33\textwidth}
        \centering
        \begin{picture}(100,100)
        \put(0,0){\includegraphics[width=1.0\linewidth]{plots_new//#4//noaa_#2_#3#5}}
        \ifnum#2=12268
            \put(32,88){#7) AR #2}
        \else
            \put(32,98){#7) AR #2}
        \fi
        \end{picture}
    \end{subfigure}
    
}
\newcommand{\jmapplots}[9]{
   \vspace{0.5cm}\hspace{-1cm}\begin{subfigure}[t]{0.4\textwidth}
        \centering
        \begin{picture}(100,100)
        \put(0,0){\includegraphics[width=0.85\linewidth,trim={0cm 1.75cm 2cm 1.5cm},clip]{plots_new//#5//noaa_#1_wb_1_#4#2_#3.pdf}}
        \put(0,99){#6)}
        \put(26,99){AR #1}
        \end{picture}
    \end{subfigure}
    \hspace{-0.8cm}\begin{subfigure}[t]{0.4\textwidth}
        \centering
        \begin{picture}(100,100)
        \put(0,0){\includegraphics[width=0.64\linewidth,trim={9cm 1.75cm 0cm 1.5cm},clip]{plots_new//#5//noaa_#1_wb_1_#8#2_#3.pdf}}
        \put(5,98){#7)}
        \end{picture}
    \end{subfigure}  
    \hspace{-3cm}\begin{subfigure}[t]{0.4\textwidth}
        \centering
        \begin{picture}(100,100)
        \put(0,0){\includegraphics[width=0.8\linewidth,trim={0cm 0cm 0cm 0cm},clip]{plots_new//#5//decay_1d_normal//noaa_#1_#3_#9.pdf}}
        \newcount\mycount 
        \mycount=`#7
        \advance\mycount by 1
        \put(3,98){\char\mycount)}
        \end{picture}
    \end{subfigure}  
    
}
\newcommand{\mgfig}[3]{
    \vspace{0.5em}
    \mgplots{12673}{12192}{#1}{#2}{#3}{a}{b}
    \mgplots{11158}{12268}{#1}{#2}{#3}{c}{d}
    \mgplots{11429}{11302}{#1}{#2}{#3}{e}{f}
    \mgplots{12297}{11339}{#1}{#2}{#3}{g}{h}
    \mgplots{11890}{11166}{#1}{#2}{#3}{i}{j}
}
\newcommand{\jmapfiga}[4]{
    \jmapplots{12673}{4}{06_0859}{#1}{#2}{a}{b}{#3}{#4}
    \jmapplots{11158}{2}{15_0135}{#1}{#2}{d}{e}{#3}{#4}
    \jmapplots{11429}{3}{06_2059}{#1}{#2}{g}{h}{#3}{#4}
    \jmapplots{12297}{1}{11_1611}{#1}{#2}{j}{k}{#3}{#4}
    \jmapplots{11890}{1}{08_0411}{#1}{#2}{m}{n}{#3}{#4}
}
\newcommand{\jmapfigb}[4]{
    \jmapplots{12192}{3}{24_2059}{#1}{#2}{a}{b}{#3}{#4}
    \jmapplots{12268}{2}{30_0023}{#1}{#2}{d}{e}{#3}{#4}
    \jmapplots{11302}{2}{26_0459}{#1}{#2}{g}{h}{#3}{#4}
    \jmapplots{11339}{6}{05_2023}{#1}{#2}{j}{k}{#3}{#4}
    \jmapplots{11166}{1}{09_2311}{#1}{#2}{m}{n}{#3}{#4}
}
\begin{document}

\title{Stability of the coronal magnetic field around large confined and eruptive solar flares}

\author{Manu Gupta\inst{\ref{inst1}} \and J. K. Thalmann\inst{\ref{inst1}}\and A. M. Veronig\inst{\ref{inst1},\ref{inst2}}}

\institute{%
University of Graz, Institute of Physics, Universit\"atsplatz 5, 8010 Graz, Austria\label{inst1} 
\email{manu.gupta@edu.uni-graz.at} \and Kanzelhöhe Observatory for Solar and Environmental Research, University of Graz, Austria\label{inst2}
}
\date{Received 22 February 2023 / Accepted 10 February 2024 }

\abstract
{
\rf{The coronal magnetic field, which overlies the current-carrying field of solar active regions, straps the magnetic configuration below. The characteristics of this overlying field are crucial in determining if a flare will be eruptive and accompanied by a coronal mass ejection (CME), or if it will remain confined without a CME.}
} 
{
\rf{In order to improve our} understanding on the pre-requisites of eruptive solar flares\rf{, we study and compare different} measures that characterize the eruptive potential of solar active regions --- the critical height for torus instability as a local measure and the helicity ratio as a global measure --- with the structural properties of the underlying magnetic field\rff{, namely the altitude of the center of the} current-carrying magnetic structure.

} 
{
Using time series of 3D \rff{optimization-based} nonlinear force-free magnetic field models \rf{for 10 different active regions (ARs)} around the time of large solar flares, we determine the altitudes of \rff{the current-weighted centers of the non-potential model structures.} Based on the potential magnetic field, we inspect the decay index, $n$, in multiple vertical planes oriented along of or perpendicular to the flare-relevant polarity inversion line, and estimate the critical height ($\hc$) for torus instability (TI) using different thresholds of $n$. The critical heights are interpreted with respect to the \rff{altitudes of the current-weighted centers of the associated non-potential structures},
as well as the eruptive character of the associated flares, and the eruptive potential of the host \rff{AR,} as characterized by the helicity ratio.
} 
{
\rff{Our most important findings are that (i) $\hc$ is more segregated in terms of flare type than the helicity ratio, and that (ii) coronal field configurations with a higher eruptive potential (in terms of the helicity ratio) also appear to be more prone to TI. Furthermore, we find no pronounced differences in the altitudes of the non-potential structures prior to confined and eruptive flares. An aspect which requires further investigation is that, quite generally, the modeled non-potential structures hardly reside in a torus-instable regime, requiring further assessment regarding  the applicability of the chosen NLFF modeling approach when targeted at the structural properties of the coronal magnetic field.}
}


\titlerunning{Stability of the coronal magnetic field around large confined and eruptive solar flares}
\maketitle

\section{Introduction}\label{sec:introduction}
Solar flares are a sudden release of a vast amount of energy stored in the coronal magnetic fields by the process of magnetic reconnection \citep[see reviews by][]{2002A&ARv..10..313P,2011SSRv..159...19F,2011LRSP....8....6S,2017LRSP...14....2B}. A coronal mass ejection (CME) is also observed with a flare when plasma and embedded magnetic field are able to successfully escape into the heliosphere \citep[for a review see, e.g.,][]{2000JGR...10523153F, 2011LRSP....8....1C}. Then the flare is called an eruptive flare. Conversely, when a flare is detected without a CME, it is referred to as a confined flare. It is a common understanding that solar eruptions such as flares and coronal mass ejections (CMEs) are different manifestations of the same physical processes causing a reconfiguration of coronal fields.

A number of possible mechanisms of storage and release type that can trigger and drive solar eruptions \mg{are known \citep[see, e.g., Table 1 in][]{2018SSRv..214...46G}. While driving mechanisms are capable of producing an eruption, trigger mechanisms can only act as an ignitor. For the latter, a driving mechanism has to take over in order to complete the eruption.} Based on these physical mechanisms, the pre-eruption magnetic configuration exist primarily in two class of models: sheared arcade or twisted flux rope \citep[for a review see, e.g.,][]{2010SSRv..151..333M,2018LRSP...15....7G}. A sheared arcade (SA) is formed by the field lines extending along the polarity inversion line (PIL) and winding less than a full turn \mgg{around the} central axis. A flux rope is formed by magnetic fluxes that wrap around a central axis above the PIL and wind with at least one turn \citep[see, e.g.,][]{2006JGRA..11112103G}. 

For a curved current channel such as magnetic flux rope (MFR), an extra force–also known as “hoop” force–arises as a consequence of the interaction between the toroidal current and the self-generated internal poloidal field \citep[e.g.,][]{1978mit..book.....B, 2016PhPl...23k2102M}.  The hoop force acts radially outwards and can be balanced by the strapping force of an external \mg{magnetic field}. However, the MFR rope may expand and eventually become unstable if the \mg{strapping force decreases} sufficiently faster than the hoop force in the direction of major \mg{radius of} the curved channel. \cite{2006PhRvL..96y5002K} referred to this ideal MHD instability as torus instability (TI). 
\cite{2018SSRv..214...46G} considered TI as one of the only two driving mechanisms, the other being the flare reconnection. \cite{aulanier_2013} suggested \mgg{breakout} model \citep[see, e.g.,][]{1999ApJ...510..485A} and TI as the only two possible mechanisms to initiate and \mg{drive solar eruptions}, and concluded all other mechanisms to be leading to, or strengthening these two mechanisms. 

\mgg{\cite{2006PhRvL..96y5002K}} investigated the stability of a partial or complete toroidal current ring by including only the hoop force and \mg{the strapping force. Using the variation of the external field perpendicular to the direction of concentrated electric current as a measure, the decay index can be quantified as a function of height. At heights where a "critical" value is exceeded, \mgg{the} TI instability may in principle act as a driver for an eruption.}
Recently it was suggested that an additional tension force which was neglected by \cite{2006PhRvL..96y5002K} may play an important role in constraining the MFR eruption, thereby resulting in a failed torus instability \citep[see, e.g.,][]{2015Natur.528..526M, 2021NatCo..12.2734Z}. \mg{Both} the hoop and tension forces in an MFR are dependent on its geometry, and therefore a range rather than a single value may  be deduced for the critical decay index (0.8 -- 1.6) based on different theoretical \mgg{\citep[e.g.,][]{2006PhRvL..96y5002K, 2010ApJ...718..433O,2010ApJ...718.1388D,2015ApJ...814..126Z}} \mg{and observation-based works} \citep[e.g.,][]{2017ApJ...843L...9W,2018ApJ...853..105B,2018ApJ...864..138J,2019ApJ...884...73D},\mg{ as well as an} experimental setup in \mgg{a} laboratory \citep{2015Natur.528..526M,2017PPCF...59a4048M,2021ApJ...908...41A}. 
\rf{Analyzing all major flares (GOES class M5 and larger) of solar cycle 24, \cite{2019ApJ...884...73D} 
found 90\% of the events to possess pre-flare MFRs much more complex than theoretical models on which TI-related thresholds are traditionally based. They suggested a lower limit of 1.3 for the critical decay index, based on the finding that all the events above this critical value exhibited an eruption. In a subsequent work, \cite{2021ApJ...907L..23D} showed that the average values of decay index decrease from before to after flares, with the value averaged over all considered flares decreasing from 0.88 to 0.63, indicative of MFRs to decrease in height through flares.} 

In recent years, TI was analyzed in a number of statistical \mgg{studies of} observed flares.
\cite{2017ApJ...843L...9W} estimated the decay index \gm{for 60 two-ribbon flares (larger than or equal to GOES class M1.0), above a segment of \mgg{the} PIL located in between the two ribbons.} They found a bimodal distribution of critical height ($\hc$) with many of the confined flares exhibiting significantly higher critical heights than the eruptive ones. Also, the decay index was found to be monotonously increasing with height for $84-86\%$ of the flares and exhibits a saddle-like profile for about $14-16\%$ of the flares. Saddle-like profiles may facilitate an eruption, nevertheless preventing the escape at upper heights, resulting in a \mg{failed eruption \citep{2022ApJ...929....2L}. \cite{2018ApJ...853..105B} calculated} the decay index for 44 flares of GOES class M5.0 or larger within vertical planes based on a linear fit to the flaring PIL. For 42  out of 44 events, they found $\hc$ larger (smaller) than 40 Mm for confined (eruptive) flares. The reason for a clear distinction among the flares in comparison to \cite{2017ApJ...843L...9W} was suspected to be different flares sizes in the two samples, with approximately $75\%$ of the events below GOES class M5.0 in \cite{2017ApJ...843L...9W} whereas all of the flares are > M5.0 in \cite{2018ApJ...853..105B}. Recently, \cite{2022A&A...665A..37J} analyzed critical heights of 42 ARs at their central \mgg{PIL}. It was observed that CME rates were $63\%$ higher during phases when critical height is increasing than when it is \mmgg{decreasing.} 

\mgg{These} stability \mgg{analyses suggest} a relationship between the critical height for TI and \mgg{whether a large flare is associated with a CME or not.} To get a full picture, however, one needs to inspect whether or not the coronal magnetic structure actually extends into the regime where TI might facilitate an eruption. Therefore it is necessary to locate \rff{the coronal altitude of the associated non-potential magnetic system} (a flux rope or sheared arcade). \mmgg{\cite{2019ApJ...877L..28Z} studied the 3D magnetic field configuration of 16 filament eruptions  that failed to be ejected and become coronal mass ejections. They found that in seven cases, the decay index at the filament apex exceed the theoretical threshold ($\nc$ = 1.5) for TI and that in all those cases a strong rotation of the filament apex was observed during the eruptions.}

\mmgg{In contrast} to the critical height for TI ---a local measure deduced from the coronal magnetic field structure--- also more global measures exist which have been found tightly related to the eruptive potential of ARs. On scales characteristic for solar ARs, one may use the helicity ratio $\hjprime$---the ratio of the helicity of the predominantly concentrated in strong electric current channels like a magnetic flux rope (MFR) or sheared arcade (SA), to the total one in the coronal volume---for details see \cite{2017A&A...601A.125P}. The helicity ratio stems from the known form of the relative magnetic helicity, $\hv$ \citep{1984GApFD..30...79B}, which can be decomposed into two components, $\hj$ and $\hpj$, the latter representing the helicity associated with the volume-threading field  \citep{2003and..book..345B}. 

The information provided by the helicity ratio has shown promise in determining the likelihood of \mgg{the CME productivity of an AR} in a more general way (since computed from volume-integrated quantities), \nrf{that is}, without making assumptions about the specific underlying coronal magnetic field configuration (e.g. a \mgg{MFR} vs a SA). \mmgg{From} simulation-based \citep[e.g.,][]{2017A&A...601A.125P,2018ApJ...863...41Z,2018ApJ...865...52L,2019A&A...628A.114P} and observation-based \citep[e.g.,][]{2018ApJ...855L..16J,2019A&A...628A..50M,2019ApJ...880L...6T} applications it became clear that the helicity ratio exhibits typically larger values for magnetic configurations\mgg{/ARs} that produce \mgg{CMEs} (compared to configurations showing none or confined activity). 
\rf{
In a first multi-event analysis, \cite{2021A&A...653A..69G} demonstrated that not only the helicity ratio, but also other AR-integrated quantities (the ratio of free to total magnetic energy and the ratio of $\hj$ to the unsigned AR flux) simultaneously exceed certain thresholds in ARs that produce eruptive flares. This finding is partially supported by the recent study of \cite{2023ApJ...945..102D}, based on a much more extended sample of flares of GOES class $\ge$M3.9 that occurred from 2011 January to 2017 December within $45\degree$ in longitude of the disk center, only disagreeing in the potential distinctive power of the helicity ratio, which they did not find distinctly different for ARs producing different types of flares. 
}
The time variation of the helicity ratio has been studied in detail by \cite{2022ApJ...937...59G} regarding its potential to indicate when key transitions occur in the coronal field \mgg{before} an eruption. They concluded, however, that further dedicated studies are required to enable a more comprehensive understanding of the (drivers of the) evolution of the coronal magnetic field and its impact on the helicity ratio.

In this \mgg{paper}, we aim to provide such a \mmgg{combined study of} a local measure for the stability of the coronal magnetic field (critical height for TI) and a global measure for the eruptive potential of an AR (relative magnetic helicity ratio) and to inspect their \mmgg{relationships to the flare type (confined or eruptive).} To do so, we combine the coronal magnetic field modeling and \mmgg{estimates of the} helicity ratios for ten solar flares \mmgg{by} \cite{2021A&A...653A..69G} (see \href{data_m}{Sect.~\ref{data_m}} paragraph 1 for a short review) with the results from a newly developed (refined) approach to compute the critical height for TI (see \href{ss:stability}{Sect.~\ref{ss:stability}} for details).

\section{Data and methods}\label{data_m}
As a {\it global} (in the sense of a characteristic active-region scale) measure for the eruptive potential of an AR, we use the helicity ratios estimated for the pre-flare corona of ten flares of GOES class M1.0 or larger that occurred within 45° of the central meridian \citep[see Table 1 in][]{2021A&A...653A..69G}. These have been based on nonlinear force-free (NLFF) coronal magnetic field time-series modeling (see Sect. 2.2 in their work for details) of selected ARs (NOAAs 12673, 11158, 11429, 12297, 11890, 12192, 12268, 11302, 11339, 11166). The time-series NLFF \mmgg{modeling was} employed in the form of two time series for each AR, one time series based on standard model parameter settings and one time series based on a setting which results in model magnetic fields with a lower residual solenoidality \mmgg{\citep[i.e., of enhanced reliability regarding the helicity computations; see][for a dedicated study]{2019ApJ...880L...6T}.} From the subsequent computation of magnetic helicities \mmgg{(the total helicity, $\hv$, as well as the helicity of the \rf{current-carrying part of the field}, $\hj$}) for each of the two NLFF models at each time \mmgg{instant, mean} values were deduced and used for further \mgg{analysis.} The final values for the helicity ratios as listed in Table 3 of \cite{2021A&A...653A..69G} then resulted from another time-averaging of pre-flare \mmgg{mean} values.

\begin{figure*}[htp]
    \captionsetup{width=\linewidth}
   \begin{subfigure}[t]{0.5\textwidth}
        \centering
        \includegraphics[width=1.0\linewidth,trim={1cm 2cm 14cm 3cm},clip]{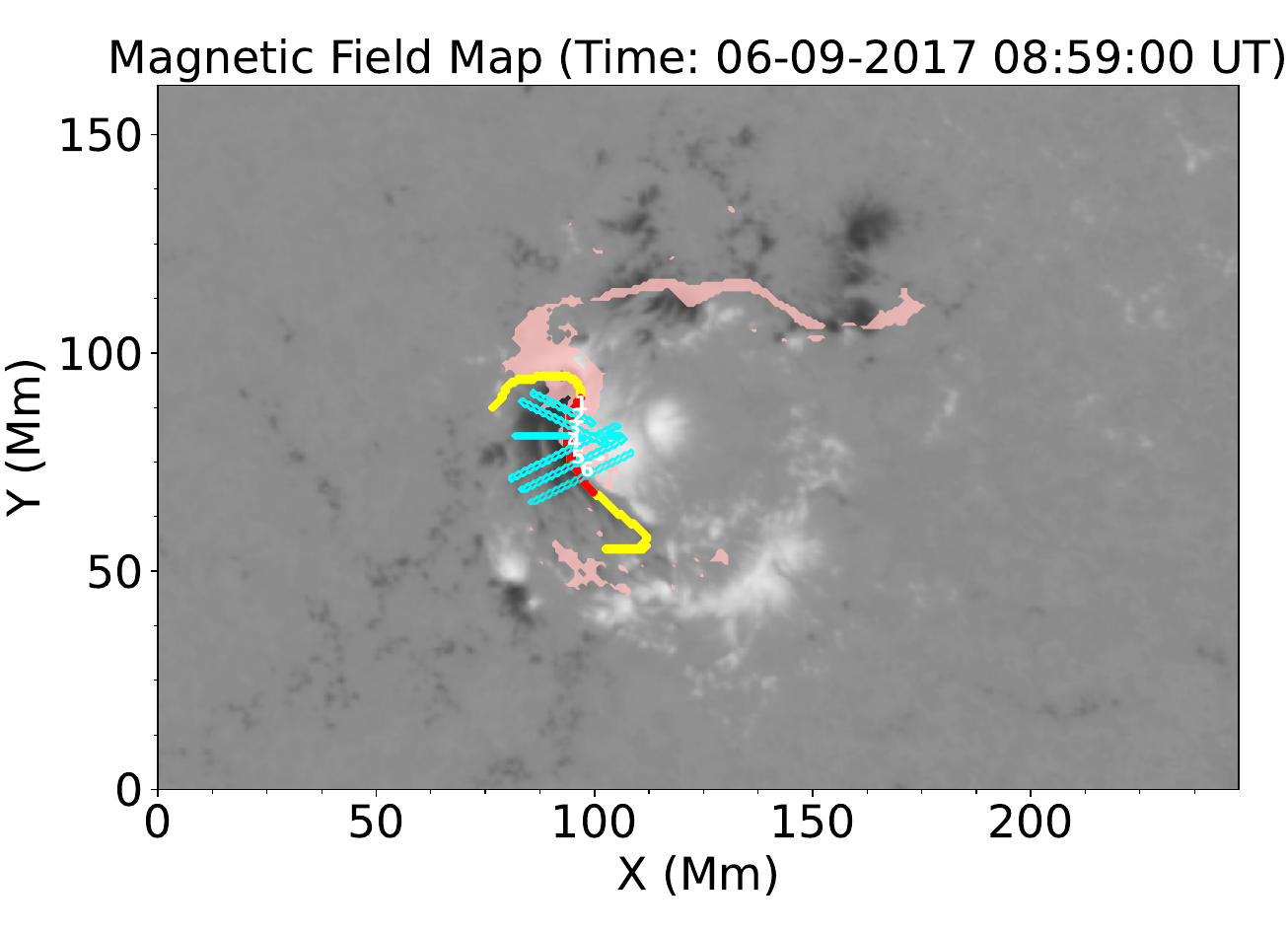}
        \put(-225,160){(a)}
    \end{subfigure}
    \begin{subfigure}[t]{0.52\textwidth}
        \centering        
        \includegraphics[width=1\linewidth,trim={0cm 0cm 0cm 0cm},clip]{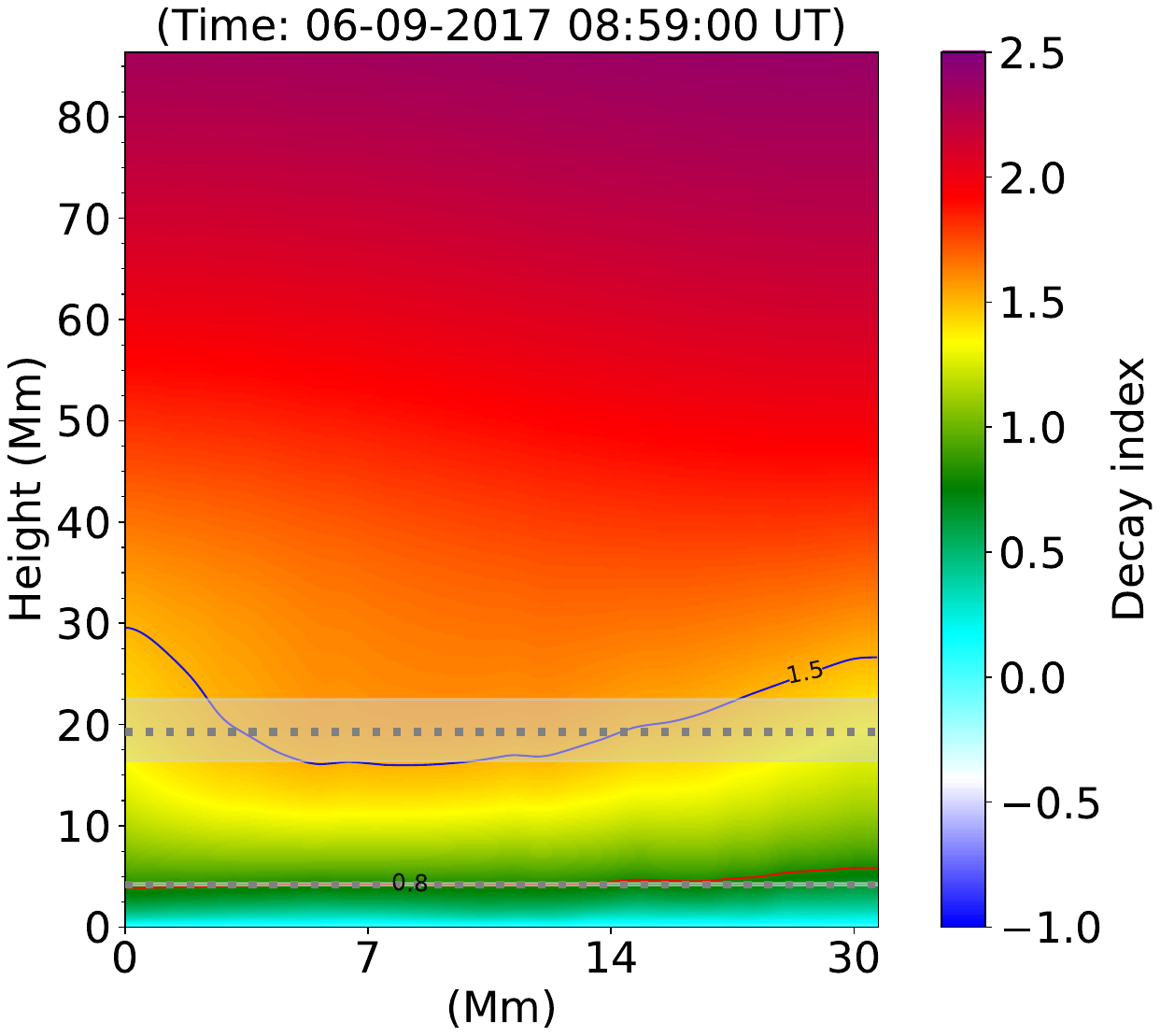}
        \put(-232,158){(b)}
    \end{subfigure}
    
    \hspace{1cm}\begin{subfigure}[t]{0.5\textwidth}
        \centering
        \includegraphics[width=1.0\linewidth]{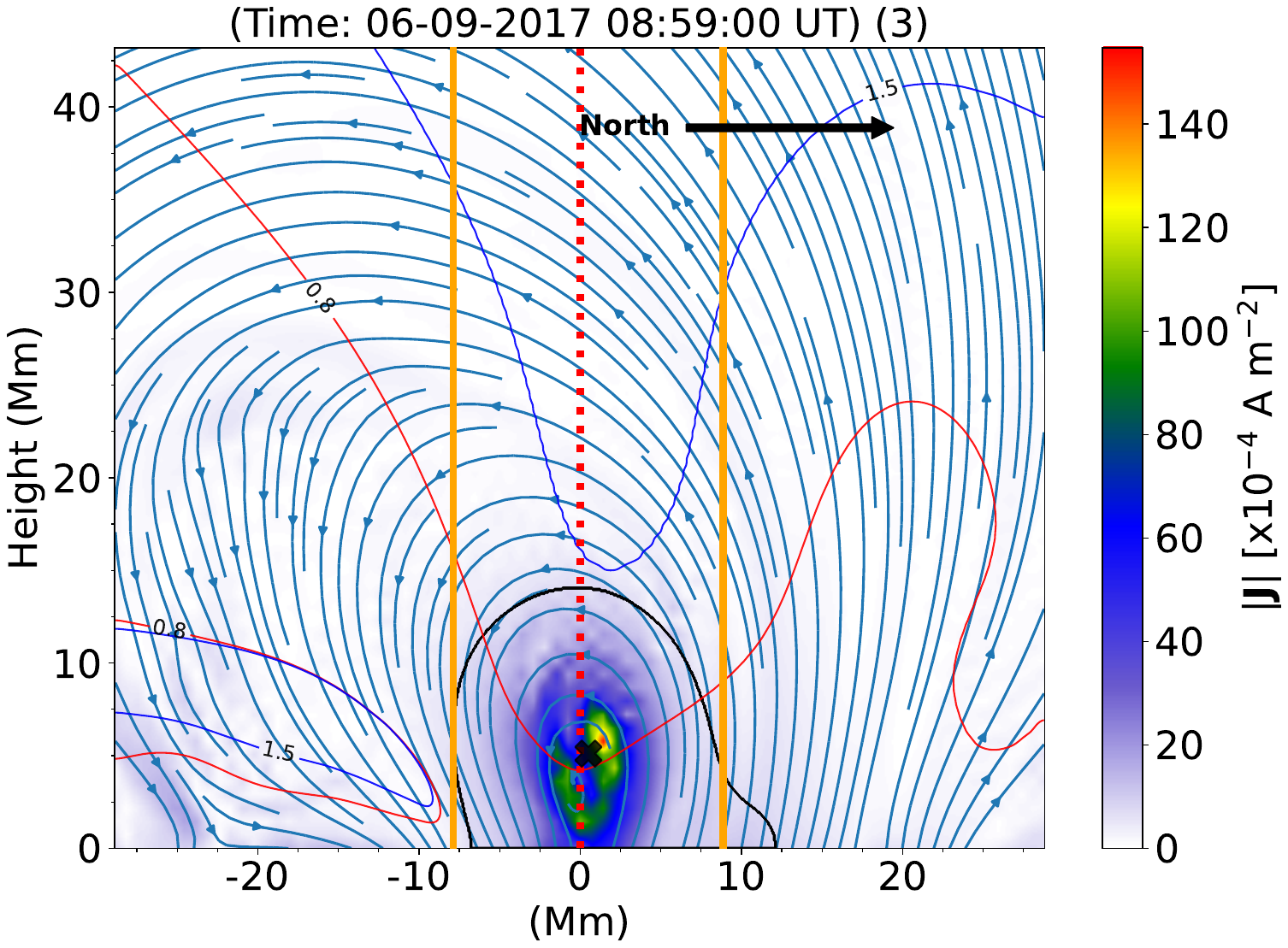}
        \put(-225,153){(c)}
    \end{subfigure}
    \hspace{-2cm}\begin{subfigure}[t]{0.4\textwidth}
        \centering
        \includegraphics[width=0.7\linewidth,trim={5cm 0cm 7cm 1cm},clip]{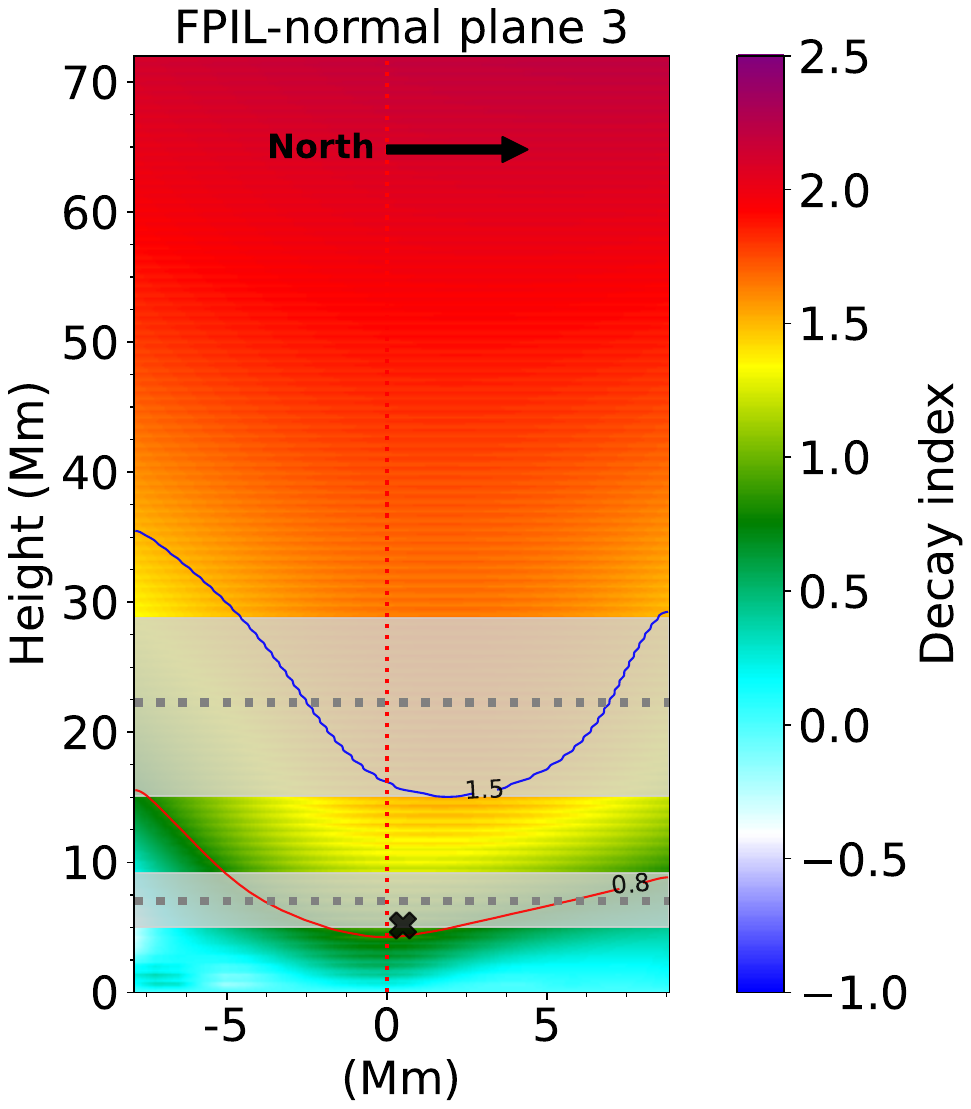}
        \put(-108,143){(d)}
    \end{subfigure}
    \hspace{-1cm}\begin{subfigure}[t]{0.1\textwidth}
        \centering
        \includegraphics[width=1.05\linewidth,trim={23cm 0cm 0cm 1cm},clip]{plots_new/noaa_12673_wb_1_norm_width_58.32Mm_decay_normal_to_pil_3_06_0859.pdf}
    \end{subfigure}\\
    \caption{
      \rf{Ingredients to the stability analysis on the example of AR 12673 on 2017-09-06 at 08:59 UT. (a) Flare ribbon mask (pink) and \rrff{central part of} flare-relevant PIL (red) along of the full detected \rrff{FPIL} of the AR (yellow curve). Cyan lines mark the footprint of the FPIL-normal planes used for further analysis. The gray-scale background shows the vertical component of the photospheric magnetic field. Black/white color denotes negative/positive polarity.
      (b) Distribution of the decay index in the FPIL-aligned vertical plane (its footprint is shown in red color in (a)). The gray dotted line indicates $\mbf{\hcmedian}$ (using $\nc$\,=\,1.5 \rff{and $\nc$\,=\,0.8)} and the shaded region mark the spread determined from the FPIL-aligned vertical plane. The blue (red) curve indicates the height where the decay index \rff{is 1.5 (0.8)}. 
      (c) Distribution of the electric current density (color-coded) and the magnetic field (blue arrows) projected into the FPIL-normal plane. The closed black contour marks the region of strongest electric current (95th percentile of \rrff{the unsigned total current density in the plane)}, used to locate the current-weighted center of MFR (SA) (black cross). \rff{The} red dotted line marks the location of the \rff{photospheric} FPIL. The vertical orange lines indicate the extent of the FPIL-plane in the horizontal direction that is considered for decay index analysis. The blue and red curves that are bending downwards indicate heights at which $\nc$ is 1.5 and 0.8, respectively.
      (d) Decay index within a selected portion (indicated by vertical orange lines in (c)) of the FPIL-normal plane. The gray dotted line indicates $\mbf{\hcbar}$ (using $\nc$\,=\,1.5 \rff{and $\nc$\,=\,0.8)} when averaged over all considered FPIL-normal planes. The gray-shaded region marks the corresponding spread of $\hc$ across all considered FPIL-normal planes. The black \rff{cross indicate} the estimated height of the \rff{current-weighted center of the} MFR (SA). The blue and red curves indicate heights at which $\nc$ is 1.5 and 0.8, respectively.}}
    \label{bz_map}
\end{figure*}

\mg{In order to \mgg{derive} a {\it local} measure for the stability of the coronal magnetic \mgg{field,} we \mmgg{apply a refined} approach to compute the critical height for \mmgg{TI.} 
\rf{
We note here that the TI as a trigger may only be relevant for a subset of our analyzed confined events, namely those that involve a failed eruption that becomes confined at larger coronal heights by a strong strapping field \citep[“type II”, following the classification of][]{2019ApJ...881..151L}. In our sample, type II confined events are the M2.0 flare in AR 12268, the M4.0 flare in AR 11302, and the M1.8 flare in AR 11339 (see Table 1). The X-class flares in our sample (hosted by AR 12192 or 11166) are type I events, \nrf{that is}, do not involve a significant eruption, and may serve as true negatives when probing their coronal configuration regarding proneness to TI.
}
This approach involves several \mgg{steps,} including the identification of the flare-relevant \mgg{PIL} and relevant information in the vertical direction in its surrounding (see \href{ss:FPIL}{Sect.~\ref{ss:FPIL}}), as well as the subsequent extraction of geometry-related (\href{ss:FPIL}{Sect.~\ref{ss:MFR}}) and \mmgg{TI\rrff{-related}} (\href{ss:stability}{Sect.~\ref{ss:stability}}) measures.}

\subsection{Flaring polarity inversion line and the corresponding \mmgg{vertical} planes}\label{ss:FPIL}
\mmgg{In order to detect the major polarity inversion line (PIL) of an AR we use the vertical magnetic field component of the  hmi.sharp\_cea\_720s data, constructed from polarization measurements of the Helioseismic and Magnetic Imager \citep[HMI;][]{2012SoPh..275..207S} on board the Solar Dynamics Observatory \citep[SDO;][]{2012SoPh..275....3P}, and used as an input to NLFF modeling in \cite{2021A&A...653A..69G}.} To identify locations of flare (ribbon) pixels, we use the SDO Atmospheric Imaging Assembly \citep[AIA;][]{2012SoPh..275...17L} \mmgg{CEA-remapped 1600 \AA\ images} and binned to the same \mmgg{plate} scale as the vector magnetic field data (720 km).

We apply the method of \cite{9377808} to detect \mgg{PILs} on a gaussian smoothed vertical magnetic field ($\Bz$) map. Secondly, to identify ribbon pixels (displayed in pink in \href{bz_map}{Fig.~\ref{bz_map}(a)}), we apply \mg{the method of} \cite{2015ApJ...801L..23T}, \mmgg{using sequences of 24-second time cadence 1600~\AA\ images covering the flare impulsive phase.} \mg{We} generate a binary map using the 99th percentile of the \mmgg{intensity of} the entire series of images as a threshold. Blooming effects are taken care of by requiring that a pixel is detected as a flaring \mgg{pixel} in at least five consecutive images. \mg{In a third step, the overlapping area of all detected PILs and the flare-pixel mask is determined, and the flare-relevant PIL \mgg{(FPIL) is} identified as that one with the largest overlap (indicated by yellow color in \href{bz_map}{Fig.~\ref{bz_map}(a)} \mgg{for AR 12673 and in Appendix \href{app1}{Fig.~\ref{app1}} for all of the ARs under study)}.}

\mg{We proceed by defining different spatial regimes in order to perform the envisaged \mmgg{stability} analysis. In particular, we define a single vertical plane along of the FPIL as one regime ("FPIL-aligned", hereafter).} Traditionally, \mg{a similar approach was used in a} number of studies to locate the heights of critical decay index \citep[see, e.g.,][]{2017ApJ...843L...9W,2018ApJ...853..105B,2018ApJ...864..138J}. \rf{The only difference to previous studies using a similar approach is that we restrict our analysis (extent of the vertical plane) to the central part (red curve in \href{bz_map}{\rrff{Fig.}~\ref{bz_map}(a)}) of the full detected FPIL (yellow curve), assuming to spot the relevant central body of the possibly unstable coronal MFR (SA) configuration.} \mmgg{As} a second spatial regime to consider, we use vertical planes, distributed equidistantly along of the \rrff{FPIL, each} of being oriented along the local FPIL-normal direction ("FPIL-normal" hereafter; \mmgg{their footprint being displayed by cyan} color lines in \href{bz_map}{\rrff{Fig.}~\ref{bz_map}(a)}). This is motivated by the fact that the geometry of \mgg{MFRs} and SAs, though often approximated in such an idealized manner, is not symmetric about the FPIL. In other words, \rff{the central axis (cross symbol} in \mgg{\href{bz_map}{Fig~\ref{bz_map}(c)})} of the involved twisted or sheared field is not necessarily located vertically above of the FPIL \mgg{(see vertical red dashed line in \href{bz_map}{Fig.~\ref{bz_map}(c)} and left column of \href{jmapa}{Fig.~\ref{jmapa}} and \href{jmapb}{~\ref{jmapb}}).} Thus, considerations of geometrical and magnetical properties in a FPIL-aligned plane may not reflect the actual conditions at those characteristic locations within MFRs and SAs. 
\mgg{These FPIL-normal \mmgg{planes enable} us to capture the projections of an \mgg{MFR}. In the next \href{ss:MFR}{Sect.~\ref{ss:MFR}} we describe the method used to locate the \rff{current-weighted center} of the MFR (SA).}
\rf{
For completeness we note that in cases where the eruption propagates in an oblique direction\rff{, the analysis} of the decay index in a direction oblique to both, the FPIL and vertical direction, might be of advantage \citep[see, e.g.,][]{2019ApJ...884...73D}. Alternatively, some authors used all pixels located at coordinates of an extended photospheric FPIL mask to study the decay index as a function of height \cite[e.g.,][]{2022A&A...665A..37J}. Exploring the differences arising from such different treatments is beyond the scope of this study, however.}

\subsection{Geometry-related measures}\label{ss:MFR}
The definition of multiple FPIL-normal planes along of the FPIL (see \href{bz_map}{Fig.~\ref{bz_map}(a)} for an example of such planes distributed along of the FPIL in AR 12673) represent a fast and easy way to investigate the distribution of magnetic field and electric currents as a function of height (see \rf{\href{bz_map}{Fig.~\ref{bz_map}(c)} for a selected plane}). To do so, we compute the components of the magnetic field and electric current normal and tangential to the FPIL-normal planes and interpolate them onto a grid with a resolution of \mmgg{72 km.} 

In order to identify the center of the MFR (SA) projected into each of the FPIL-normal planes, we assume the central region to be represented in the form of \mgg{a channel of} strongest concentrated electric current. We use the 95th percentile of the unsigned electric current density in each of the FPIL-normal planes as selection \mgg{criterion. Hereafter}, we use a map of the Gaussian-smoothed total current density to find the corresponding current-weighted center. We show an exemplary result \rf{for the NLFF model} of AR 12673 at 2017-Sep-06 08:59~UT in \href{bz_map}{Fig.~\ref{bz_map}(c)} where the detected \rrff{center} of the strong current concentration present \rrff{is} marked by \rrff{the black cross}. 
The corresponding height (altitude in the model corona) is referred to as $\hcenter$ hereafter. For further analysis, we compute the mean value $\hcentermean$ and standard deviation from the range of values deduced from all of the FPIL-normal planes for each time step. \rrff{We remark here that the applied averaging lowers the final measure for the altitude of the non-potential structure ($\hcenter$), in comparison to the altitude of the apex of the non-potential structure determined from a single plane which is used in many studies in the literature (see the dedicated discussion in \href{Discussion}{Sect.~\ref{Discussion}}).} We note here that in contrast to \cite{2021A&A...653A..69G}, we employ only one NLFF model with the standard model parameter settings, as both NLFF models from \cite{2021A&A...653A..69G} generate similar estimates of $\hcenter$ of the \rrff{current channel}.

\subsection{Stability analysis}\label{ss:stability}
\mg{Following \cite{2006PhRvL..96y5002K}, the decay index is defined as,}
\begin{eqnarray}
        n&=&-\frac{d(ln(B_{\mathrm{hor}}))}{d(ln(h))},
\end{eqnarray}
where $B_\mathrm{{hor}}$ represents the strapping field approximated by the horizontal component of the potential magnetic field and $h$ is the height in the corona. \rf{A larger value of $n$ indicates a potential field to more rapidly decay with height.} The \rf{TI} 
may occur in regions where $B_\mathrm{{hor}}$ declines with height at a sufficiently steep rate.

The potential field is computed by applying a Green's function method to the map of $\Bz$ used for PIL detection. We compute the decay index from the potential field projected into each of the FPIL-normal planes \rf{(see \rf{\href{bz_map}{Fig.~\ref{bz_map}(d)} for a selected plane}).}

We restrict the computation of the decay index to the portion of the FPIL-normal planes that cover the relevant non-potential core field, as indicated exemplary in \href{bz_map}{Fig.~\ref{bz_map}(c)} by the orange vertical lines.
\rrff{This specific section is determined by the width between the center and the edge of the current density contour (e.g., black contour in \href{bz_map}{Fig.~\ref{bz_map}(c)}), which is smaller on \nrf{one side}, and then reciprocated on the opposite side.}
From \rf{\href{bz_map}{Fig.~\ref{bz_map}}(d)} it is evident that the decay index can vary significantly within short distances in a direction perpendicular to the \rff{PIL}. Designed as is, our method is sensitive to those variations and, computing the critical height for torus instability from the such defined in-plane distribution of $n\,(h)$, the correspondingly derived values for $\hc$ too. \rrff{We compute the median ($\mbf{\nmedian}$) and median absolute deviation ($MAD$) of the decay index in the horizontal direction across all FPIL-normal planes}, covering an area of the width \rrff{for each FPIL-normal plane} as described \rff{above on} both sides of the \rff{current-weighted center} (marked by orange vertical lines in \href{bz_map}{Fig.~\ref{bz_map}(c)}). \rrff{We remark here that this averaging of the decay index values in the FPIL-normal planes tends to deliver $\hcmedian$ values that are higher than the $\hc$ values at the position of the current-weighted center of the non-potential structure (compare the run of the red curve and the position of the black cross in \href{bz_map}{Fig.~\ref{bz_map}(c)}).}

\begin{figure*}[ht]
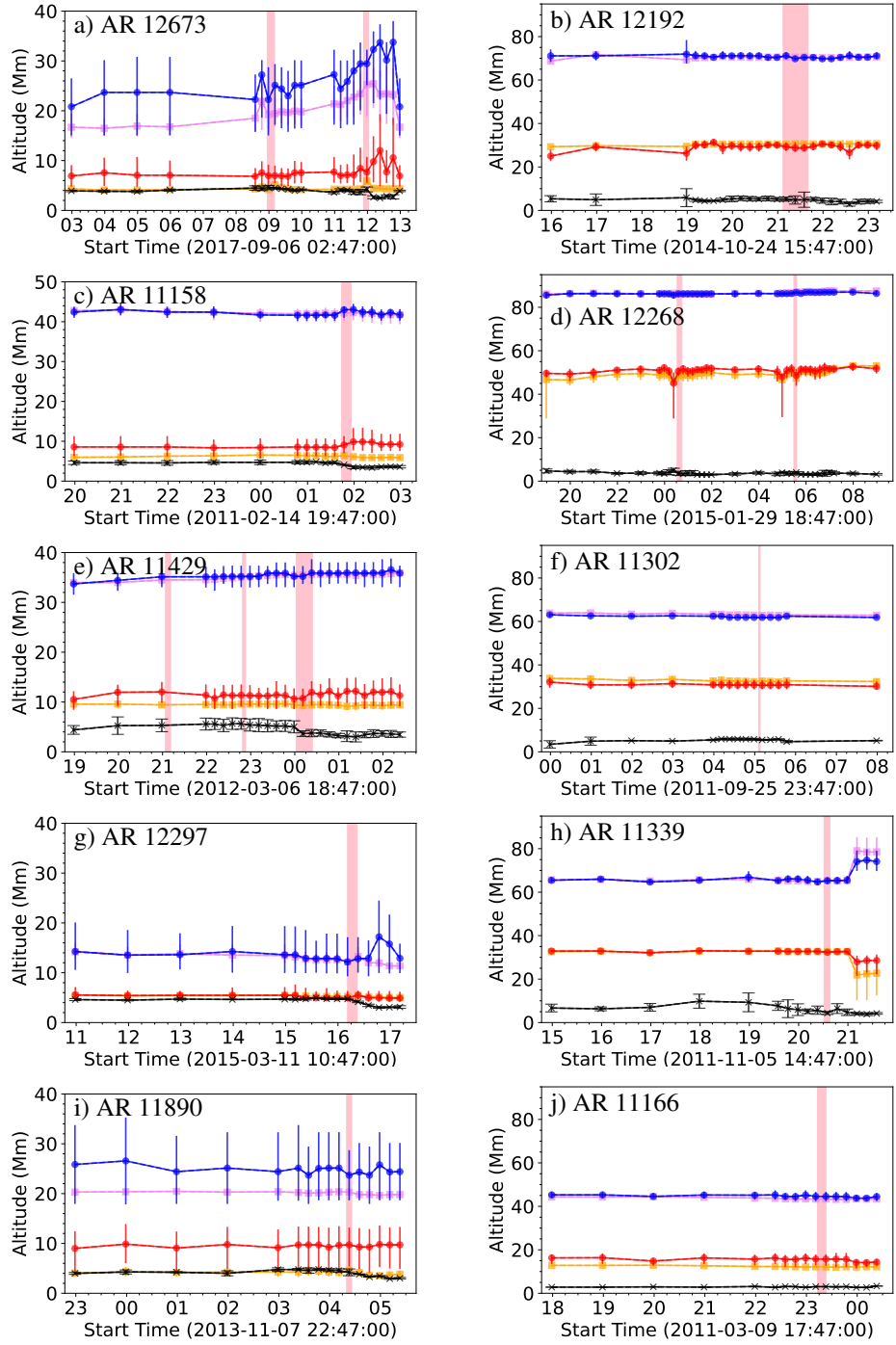

    \captionsetup{width=\linewidth}
    \mgfig{width_58.32Mm_critical_height_decay_median_proj_ht_center_mean_wb_1}{hcrit_decay_median}{.pdf}
    \vspace{-1em}
    \caption{\rf{Time evolution of critical heights for TI, $\mbf{\hcbar}$ and $\mbf{\hcmedian}$, (using two different values for $\nc$), as well as the \rff{center ($\mbf{\hcentermean}$) heights} around the five eruptive (left column) and five confined (right column) flares under study. \rff{The solid black lines connecting crosses mark the values of $\mbf{\hcentermean}$.} Blue and red lines represent $\mbf{\hcbar}$ when $\nc$\,=\,1.5 or $\nc=0.8$, respectively, is used as a threshold for TI, computed from the FPIL-normal planes. Violet and orange lines, respectively, show $\mbf{\hcmedian}$ computed from the FPIL-aligned planes. \rff{Vertical bars mark the corresponding uncertainty.} }
    }
    \label{$hcmedian$_main}
\end{figure*}

We are then able to compute the critical height for TI 
\rf{assuming different values of $n$ to be critical (where we use once $\nc$\,=\,1.5 as commonly done in related analyses, and once $\nc=0.8$ ranging at the lower end of possibilities \citep[e.g.,][]{2016PhPl...23k2102M}.} 
Uncertainties are estimated \rf{based on the variations} 
($\mbf{\nmedian}\pm1MAD$). \rrff{The analysis then includes one further step, which is the averaging of the $\nbf{\nmedian (h)}$ profiles over all FPIL-normal planes, leading to a single value $\mbf{\nbar}$ at a given height and the critical height $\mbf{\hcbar}$ (i.e., $\mbf{h\,(\overline{n}=n_\mathrm{crit})}$ in \href{$hcmedian$_main}{Fig.~\ref{$hcmedian$_main}}.).} In case of saddle-like profiles this condition can cause \rf{complications} 
when \rrff{the uncertainty in $\mbf{\hcbar}$ based on the profiles of $\mbf{\nbar}+1MAD$ and $\mbf{\nbar}-1MAD$ is not simultaneously located above and below the $\mbf{\hcbar}$.} In such cases we select the smaller value of variation to represent the uncertainty in both directions. \rrff{Similarly, we estimate $\hcmedian$ and $MAD$ from the FPIL-aligned plane (indicated by gray dotted lines and shaded regions in \href{bz_map}{Fig.~\ref{bz_map}(b)} for $\nc$\,=\,1.5 and $\nc$\,=\,0.8). We remark here explicitly again that the resulting values of $\hc$ that correspond to the FPIL-aligned planes result from the averaging along of one direction (along of each FPIL-aligned plane) while those based on the analysis of sets of FPIL-normal planes are based on the averaging along of two directions (once along each FPIL-normal plane and once over all FPIL-normal planes, \nrf{that is}, basically along of the FPIL).}

\rrff{From the time profiles of the quantities introduced, that is, $\hcentermean$ and $\hcbar$ (in \href{$hcmedian$_main}{Fig.~\ref{$hcmedian$_main}}), we estimate a "characteristic" pre-flare level (\href{t1}{Table~\ref{t1}}) as done by \cite{2021A&A...653A..69G}. This involves taking all data points within a five-hour window prior to the flare onset, and compute the mean value (hereafter referred to as "time-averaged" and denoted by angular brackets) and the standard deviation.}

\section{Results}
\rf{In the following, for each of the ARs under study, we describe the flares produced during the \mgg{time series under study}, and examine the magnitudes and the evolution of the \rff{MFR (SA) current-weighted center} and the critical height. In \href{$hcmedian$_main}{\rrff{Fig.}~\ref{$hcmedian$_main}}, we show the time evolution of \rff{$\hcentermean$, $\mbf{\hcbar}$ and $\mbf{\hcmedian}$} (twofold, for using $\nc=0.8$ or $\nc$\,=\,1.5 as a threshold) computed from both, the FPIL-normal and FPIL-aligned planes. In \href{t1}{Table~\ref{t1}}, we summarize the derived time-averaged pre-flare values of $\mbf{\hcbar}$ computed from the FPIL-normal planes (since we believe that this measure, due to the usage of multiple FPIL-normal planes, is more robust than the estimate based on a single plane only) \rff{and $\hcentermean$,} where we use the time instances during last five hours before the flare start.}

In the Appendix, we show the equivalent of \rrff{\href{bz_map}{\rrff{Fig.}~\ref{bz_map}}} for each of the ARs at an instance before the time of large flares. In \href{app1}{Fig.~\ref{app1}}, we display the detected FPIL, flare mask, and the FPIL-normal planes on top of the vertical component of the photospheric magnetic field. For the same instance, \href{app1_1}{Fig.~\ref{app1_1}} and \href{app1_2}{~\ref{app1_2}} show the FPIL-aligned planes (left panels) and $\nbf{\nmedian (h)}$ profiles (right panels). In \href{jmapa}{Fig.~\ref{jmapa}} and \href{jmapb}{~\ref{jmapb}}, we display the total current density and the projected NLFF field (computed based on the standard setting for free model parameters) on selected FPIL-normal planes (in first column), decay index map (in second column), and $\mbf{\nbar}$ vs $h$ curves derived from all of the FPIL-normal planes (in last column). \rff{We note that, for each of the ARs that produced confined flares (NOAAs 12192, 12268, 11302, 11339, 11166), the NLFF model indicates the presence of MFRs except in case of AR 12192 where only a SA is recovered from NLFF modeling. These coronal settings are consistent with those in earlier studies.} 

\begin{table*}[ht]
\caption{NOAA number, GOES SXR flare class, time-averaged pre-flare values of critical 
heights (using $\nc=0.8$ or $\nc$\,=\,1.5), \rff{current-weighted center} of MFR (SA) derived from FPIL-normal planes, and 
\rff{height differences computed from the values of $\hcang^{(\nc=0.8)}$ and $\hcentermean$ (in the second last column)} 
for the flares under study. Time averages are computed from all data points within a five-hour time window prior to the nominal flare start time. \rf{The first five flares in the list were eruptive flares.}
}
\label{t1}
\centering
\begin{tabular}{lcccccccc}
\hline\hline
\rule{0pt}{4ex}NOAA & SXR class &\makecell{\mrf{$\mbf{\hcang}^{(n=0.8)}$}} &$\langle\hcentermean\rangle$ &\makecell{\mrf{$\mbf{\hcang}^{(n=0.8)}-\langle\hcentermean\rangle$}} &\makecell{$\mbf{\hcang}^{(n=1.5)}$}\\
\hline
 12673 &X9.3& $7.28\pm0.44$&  $4.12\pm0.32$& $3.16\pm0.77$&  $25.34\pm2.12$\\
 11158 &X2.2& $8.69\pm0.24$&  $4.46\pm0.17$& $4.23\pm0.41$& $42.07\pm0.59$\\
 11429 &X5.4& $11.34\pm0.37$& $5.36\pm0.19$& $5.98\pm0.56$&  $34.28\pm0.38$\\
 12297 &X2.1& $5.24\pm0.24$&  $4.70\pm0.11$& $0.54\pm0.36$&  $13.34\pm0.48$\\
 11890 &X1.1& $9.54\pm0.29$&    $4.51\pm0.26$& $5.03\pm0.55$&  $24.96\pm0.74$\\
\hline
 12192 &X3.1& $29.42\pm1.16$& $5.10\pm0.44$& $24.32\pm1.60$& $71.05\pm0.41$\\
 12268 &M2.0& $50.11\pm2.08$& $4.02\pm0.48$& $46.09\pm2.56$& $86.13\pm0.20$\\
 11302 &M4.0& $30.85\pm0.2$& $5.52\pm0.39$& $25.33\pm$0.59& $62.20\pm0.32$\\
 11339 &M1.8& $32.75\pm0.24$& $7.04\pm1.52$& $25.72\pm$1.77& $65.63\pm0.68$\\
 11166 &X1.5& $15.82\pm0.47$& $2.92\pm0.14$& $12.90\pm$0.618& $44.93\pm0.34$\\
\hline
\end{tabular}
\end{table*}

\subsection{NOAA AR 12673}\label{AR12673}
AR 12673 hosted two X-class flares on 2017 September 6. The first X2.2 flare was observed as a confined flare and 2.6 hours later X9.3 flare was \mgg{observed along with a fast CME}. \cite{2020ApJ...890...10Z} analyzed the evolution of the MFR around the time of these events by combining EUV observations with the magnetic field extrapolation. \mgg{In their study, the MFR height was tracked by first calculating the twist number on a vertical plane and then assuming the boundaries of the MFR where the twist number is equal to -1.} They found that the MFR reached high enough to activate a null point, but was still below the threshold of torus instability and therefore could not erupt successfully. However, after the first flare the MFR continuously expanded while forcing concurrent null point reconnections thereby weakening the overlying field. Eventually the MFR reached a critical height and became torus unstable. \rff{Here, we find that the $\hcentermean$ estimated based on \rrff{the current density in the} FPIL-normal planes varies between \mrf{$2.5 \pm 0.2$} Mm and \mrf{$4.6 \pm0.4$} Mm during the time series.} The critical height, \rf{$\mbf{\hcbar^{(n=1.5)}}$}, on the FPIL-normal planes 
\rrff{varies} from 
\mrf{$20.8$} Mm to \mrf{$33.8$} Mm and $\mbf{\hcmedian^{(n=1.5)}}$ \rf{\rrff{varies} from} \mrf{$16.5$} Mm to \mrf{$25.4$} Mm \rf{on the FPIL-aligned plane.} 
\rf{The $\mbf{\hcbar^{(n=0.8)}}$ estimated from the FPIL-normal planes ranges from $6.8$ Mm to $12$ Mm, and $\mbf{\hcmedian^{(n=0.8)}}$ \rrff{estimated from the FPIL-aligned plane} ranges from $4.1$ Mm to $5.9$ Mm.}
\subsection{NOAA AR 11158}\label{AR11158}
The largest flare \mgg{that} originated from AR 11158 was an X2.2 (SOL2011-02-15T01:44) eruptive flare.  Tether-cutter reconnection was suggested to be the most likely trigger of eruption for this flare (\cite{2012ApJ...760...31K,2012ApJ...748L...6N,2013ApJ...770...79I}). Furthermore, the temporal evolution of the twisted field lines was found to play an important role in various confined and eruptive flares produced by \mgg{this} AR. \rf{\cite{2011ApJ...741L..35Z} suggested the presence of an MFR from observation of a sigmoid and an erupting loop-like feature before and during the eruption. Based on their analysis, as the magnetic configuration evolves, the MFR reaches a stage where it becomes unstable and loses equilibrium, leading to a flare reconnection under the MFR.}

In the time series \mgg{(\href{$hcmedian$_main}{Fig.~\ref{$hcmedian$_main}}(c))}, \rff{the $\hcentermean$ varies in the approximate range between \mrf{$3.4 \pm0.1$} Mm and \mrf{$4.7\pm0.1$} Mm.} The 
\mrf{$\mbf{\hcbar^{(n=1.5)}}$} on FPIL-normal planes \mgg{is between} \mrf{$41.6$} Mm and \mrf{$43.1$} Mm, and $\mbf{\hcmedian^{(n=1.5)}}$ on the FPIL-aligned \mgg{plane from} \mrf{$41.7$} Mm to \mrf{$43.0$} \mgg{Mm.} A similar value of the critical height ($\approx42$ Mm) was reported by \cite{2015ApJ...804L..28S} and \cite{2018ApJ...860...58V} for the X2.2 flare event based on the decay index above the flaring PIL. \rf{The $\mbf{\hcbar^{(n=0.8)}}$ estimated from the FPIL-normal planes ranges from $8.4$ Mm to $9.7$ Mm, and $\mbf{\hcmedian^{(n=0.8)}}$ \rrff{estimated from the FPIL-aligned plane} ranges from $5.9$ Mm to $6.6$ Mm.}

\subsection{NOAA AR 11429}\label{AR11429}
In the time series of AR 11429 (\href{$hcmedian$_main}{Fig.~\ref{$hcmedian$_main}}(e)), the X5.4 flare (SOL2012-03-07T00:02) is eruptive and two M-class flares are confined. \rff{The values of $\hcentermean$ vary between \mrf{$3\pm1$} Mm and \mrf{$5.7\pm1.2$} Mm (\href{$hcmedian$_main}{Fig.~\ref{$hcmedian$_main}}(e)).} During the time series, 
\mrf{$\mbf{\hcbar^{(n=1.5)}}$} estimated on \gmnew{FPIL-normal planes} spans \mgg{in the} range of \mrf{$33.7$} Mm to \mrf{$36.6$} Mm, and $\mbf{\hcmedian^{(n=1.5)}}$ \rrff{estimated} on FPIL-aligned plane spans from \mrf{$34$} Mm to \mrf{$35.9$} Mm. A similar value of the critical height ($\approx34$ Mm) has been reported by \cite{2015ApJ...804L..28S} for the X5.4 flare event. \rf{The $\mbf{\hcbar^{(n=0.8)}}$ estimated from the FPIL-normal planes ranges from $10.5$ Mm to $12.2$ Mm, and $\mbf{\hcmedian^{(n=0.8)}}$ \rrff{estimated from the FPIL-aligned plane} ranges from $9.1$ Mm to $9.6$ Mm.}

\subsection{NOAA AR 12297}\label{AR12297}
AR 12297 produced 17 M-class and an X2.1 flare (SOL2015-03-11T16:11), and most of them were eruptive. The reason for that may be the low critical heights which are the lowest in our sample with a value of $\approx12$ Mm before the X2.1 flare. Before the flare, \cite{2016ApJ...830..152L} observed a slipping motion at the north part of the MFR that lasted about 40 minutes and successfully peeled off the flux rope. Eventually, the flux rope started to rise up and the X2.1 flare was initiated triggered by the magnetic reconnection observed at the top boundary of the MFR. In \href{$hcmedian$_main}{Fig.~\ref{$hcmedian$_main}}(g), \rff{\mgg{$\hcentermean$} during the time series varies between \mrf{$3\pm0.2$} Mm and \mrf{$4.9$} Mm.}
The critical height \rf{$\mbf{\hcbar^{(n=1.5)}}$ estimated from the} 
\gmnew{FPIL-normal planes} during the time series spans from \mmgg{$12.2$ Mm to $17.2$ Mm}, and on \gmnew{FPIL-aligned} plane $\mbf{\hcmedian^{(n=1.5)}}$ spans from \mrf{$11.4$} Mm to \mrf{$14.2$ Mm}. \mmg{A saddle like profile is also observed in the decay index (see fourth row in \href{app1_1}{Fig.~\ref{app1_1}} and \href{jmapa}{Fig.~\ref{jmapa}}).}
\rf{The $\mbf{\hcbar^{(n=0.8)}}$ estimated from the FPIL-normal planes ranges from $4.9$ Mm to $5.5$ Mm, and $\mbf{\hcmedian^{(n=0.8)}}$ \rrff{estimated using the FPIL-aligned plane} ranges from $5$ Mm to $5.5$ Mm.}

\subsection{NOAA AR 11890}\label{AR11890}
AR 11890 produced an X1.1 eruptive flare (SOL2013-11-08T04:20). The corona of AR 11890 above the FPIL constantly exhibits a low critical height. During the time series (\href{$hcmedian$_main}{Fig.~\ref{$hcmedian$_main}}(i)), \rff{\mgg{$\hcentermean$} varies between \mrf{$2.9\pm0.1$} Mm and \mrf{$4.8\pm0.2$} Mm.} 
\rf{The FPIL-normal based $\mbf{\hcbar^{(n=1.5)}}$} ranges from \mrf{$23.7$} Mm to \mrf{$26.6$ Mm}, and $\mbf{\hcmedian^{(n=1.5)}}$ on the FPIL-aligned plane ranges between $19.7$ Mm and $20.4$ Mm. \mmg{In addition, a saddle like profile is observed in the decay index with saddle bottom at a large height ($\approx70$ Mm; see last row in \href{app1_1}{Fig.~\ref{app1_1}} and \href{jmapa}{Fig.~\ref{jmapa}}).} \rf{The $\mbf{\hcbar^{(n=0.8)}}$ estimated from the FPIL-normal planes ranges from $9$ Mm to $9.9$ Mm, and $\mbf{\hcmedian^{(n=0.8)}}$ \rrff{estimated using the FPIL-aligned plane} ranges from $3.5$ Mm to $4.6$ Mm.}

\subsection{NOAA AR 12192}\label{AR12192}
AR 12192 produced \mgg{numerous} M- and X-class flares, all except an M4.0 flare confined. The largest flare that is being investigated here was an X3.1 flare (SOL2014-10-24T21:07).  \cite{2015ApJ...801L..23T} and \cite{2015ApJ...804L..28S} investigated this AR and found strong overlying fields that prevented the core fields from erupting. During the time series (\href{$hcmedian$_main}{Fig.~\ref{$hcmedian$_main}}(b)), \mrf{$\mbf{\hcbar^{(n=1.5)}}$ ranges} 
from \mrf{$69.8$} Mm to \mrf{$71.9$} Mm \rf{as estimated from the} 
FPIL-normal planes and $\mbf{\hcmedian^{(n=1.5)}}$ from \mrf{$68.8$} Mm to \mrf{$71.7$} \rf{using the} 
the FPIL-aligned plane. Similar values of the critical height were reported by \cite{2015ApJ...804L..28S}, \cite{2017ApJ...843L...9W}, and \cite{2018ApJ...853..105B} for the X3.1 flare event. \rff{The estimated height of the current-weighted center of the sheared arcade (an MFR could not be detected) in this case ranges between \mrf{$3 \pm 1$} Mm and \mrf{$5.9 \pm 4.1$} Mm.} \mgg{\cite{2016ApJ...828...62J} studied the physical mechanism of the X3.1 flare using a data-driven MHD model, and found a tether-cutting reconnection between the sheared magnetic arcades as the trigger mechanism.} \rf{The $\mbf{\hcbar^{(n=0.8)}}$ estimated from the FPIL-normal planes ranges from $25$ Mm to $31.3$ Mm, and $\mbf{\hcmedian^{(n=0.8)}}$ \rrff{estimated from the FPIL-aligned plane} ranges from $29.3$ Mm to $30.7$ Mm.}

\subsection{NOAA AR 12268}\label{AR12268}
AR 12268 produced only confined flares \citep{2019ApJ...871..105Z}. In this study, we analyze an M2.0 (SOL2015-01-30T00:32) and an M1.7 (SOL2015-01-30T05:29) \mgg{confined flare from this AR.} \cite{2021NatCo..12.2734Z} studied the M2.0 flare event in great detail by performing a data-driven-magnetohydrodynamic simulation. They found that a component of the Lorentz force, resulting from the radial magnetic field of the MFR played a major role in constraining the eruption. In their study, a failed torus regime was observed, followed by a failed kink regime \citep[see Fig. 8 of][]{2021NatCo..12.2734Z}. 
The \rf{pre-flare} average value of \mrf{$\mbf{\hcbar^{(n=1.5)}}$} 
is \gmnew{$\approx86$ Mm} which is much higher than the \rff{altitude of the current-weighted center} of the MFR (\href{$hcmedian$_main}{Fig.~\ref{$hcmedian$_main}}(d)). The critical height observed here is the highest in our set of ARs. During the time series (\href{$hcmedian$_main}{Fig.~\ref{$hcmedian$_main}}(d)), \rff{\mgg{$\hcentermean$} varies between \mrf{$2.9\pm0.2$} Mm and \mrf{$4.9 \pm1$} Mm.} 
The \mrf{$\mbf{\hcbar^{(n=1.5)}}$ estimated from the} 
FPIL-normal planes ranges from \mrf{$85.6$ Mm} to \mrf{$87$} Mm, and 
$\mbf{\hcmedian^{(n=1.5)}}$ \rrff{estimated from the FPIL-aligned plane} ranges from \mrf{$86.1$} Mm to \mrf{$87.6$ Mm}. \rf{The $\mbf{\hcbar^{(n=0.8)}}$ estimated from the FPIL-normal planes ranges from $45.1$ Mm to $52.7$ Mm, and $\mbf{\hcmedian^{(n=0.8)}}$ \rrff{estimated from the FPIL-aligned plane} ranges from $45.8$ Mm to $53.1$ Mm.} \rff{Here also, a saddle like profile is observed in the decay index (see second row in \href{app1_2}{Fig.~\ref{app1_2}} and \href{jmapb}{Fig.~\ref{jmapb}}).}

\subsection{NOAA AR 11302}\label{AR11302}
In \rff{the} case of AR 11302, we analyze the evolution of the corona around the M4.0 confined flare (SOL2011-09-26T05:06). The failed eruption was analyzed by \cite{2022ApJ...926..143M}, \mgg{who} 
\rf{
found a saddle-like profile with a local decrease of the decay index above $\nc$\,=\,1.5 at very low heights in the model atmosphere ($\sim$3\,Mm). In sharp contrast, we find a $\nbf{\hcbar^{(n=1.5)}}\gtrsim60$\,Mm, which we consider as plausible for the following reasons. \cite{2022ApJ...926..143M} used a single vertical plane in their analysis, located along one of the two identified flux ropes where the flare was observed. If we would consider a similarly positioned plane only, we would find a similar distribution of $n(h)$ and estimate for $\hc$. However, we are using multiple FPIL-normal planes in our analysis, and for an even only slightly displaced FPIL-normal plane the distribution of $n(h)$ looks very different (see 3rd row in \nrf{\href{app1_2}{Fig.~\ref{app1_2}}}). As a consequence, our $\mbf{\nbar}$ distribution after averaging over all FPIL-normal planes does deliver $\nbf{\hcbar^{(n=1.5)}}$ at distinctly larger heights ($\gtrsim$60\,Mm as shown in the right panel in 3rd \nrf{row in \href{jmapb}{Fig.~\ref{jmapb}}}). In favor of our estimate of $\nbf{\hcbar^{(n=1.5)}}$, we argue that the value of the decay index around a low-lying null point cannot be regarded as being representative for the impact of the large-scale strapping (potentially confining) \rrff{field. This} is also supported from our FPIL-aligned estimate of $\hc$, residing at $\gtrsim$60\,Mm.
}
The \rf{FPIL-normal based estimate of $\mbf{\hcbar^{(n=1.5)}}$ ranges} 
from \mrf{$61.8$} Mm to \mrf{$63.1$} Mm, and \rf{when based} on the FPIL-aligned plane $\mbf{\hcmedian^{(n=1.5)}}$ \rf{ranges} from \mrf{$62.8$} Mm to \mrf{$63.9$} Mm. 
\rff{The \mgg{$\hcentermean$} varies between $3.4\pm1.7$ Mm to $5.9\pm0.4$ Mm.} \rf{The $\mbf{\hcbar^{(n=0.8)}}$ estimated from the FPIL-normal planes ranges from \mrf{$30.1$} Mm to \mrf{$32.1$} Mm, and $\mbf{\hcmedian^{(n=0.8)}}$ \rrff{estimated from the FPIL-aligned plane} ranges from $32.2$ Mm to $33.8$ Mm.}

\subsection{NOAA AR 11339}\label{AR11339}
\rff{In the case of AR 11339, we analyze the evolution of corona around the} confined M1.8 flare (SOL2011-11-05T20:31). During the time series (\href{$hcmedian$_main}{Fig.~\ref{$hcmedian$_main}}(h)), \rff{\mgg{$\hcentermean$} spans between $3.9\pm0.7$ Mm and $9.8\pm3.2$ Mm.} The \rf{$\mbf{\hcbar^{(n=1.5)}}$} 
on \gmnew{FPIL-normal planes ranges from \mrf{$64.7$} Mm to \mrf{$74.7$} Mm, and $\mbf{\hcmedian^{(n=1.5)}}$ on the FPIL-aligned plane ranges from \mrf{$64.9$} Mm to \mrf{$79.1$} Mm}. \rf{The $\mbf{\hcbar^{(n=0.8)}}$ estimated from the FPIL-normal planes ranges from \mrf{$27.9$} Mm to \mrf{$33$} Mm, and $\mbf{\hcmedian^{(n=0.8)}}$ \rrff{estimated from the FPIL-aligned plane} ranges from \mrf{$21.9$} Mm to \mrf{$33$} Mm.}

\subsection{NOAA AR 11166}\label{AR11166}
In the case of AR 11166, we analyze the evolution of corona around the confined X1.5 flare (SOL2011-03-09T23:13). 
During the time series (\href{$hcmedian$_main}{Fig.~\ref{$hcmedian$_main}}(j)), \rff{the \mgg{$\hcentermean$} span between \mrf{$2.7$} Mm to \mrf{$3.3$} Mm}. The \mrf{$\mbf{\hcbar^{(n=1.5)}}$} 
on FPIL-normal planes ranges from \mrf{$43.8$} Mm to \mrf{$45.3$} Mm, and $\mbf{\hcmedian^{(n=1.5)}}$ on the FPIL-aligned plane spans between \mrf{$43.2$} Mm to \mrf{$44.4$} Mm. \rf{The $\mbf{\hcbar^{(n=0.8)}}$ estimated from the FPIL-normal planes ranges from $14.1$ Mm to $16.3$ Mm, and $\mbf{\hcmedian^{(n=0.8)}}$ \rrff{estimated from the FPIL-aligned plane} ranges from $12$ Mm to $12.9$ Mm.}

\section{Discussion}\label{Discussion}

In this study, we calculated the critical height \mmgg{for} torus instability and estimated the approximate \rff{coronal altitude of the current-weighted center of \rrff{the} MFR (SA).} 
\mgg{To do so, we identified} the flare-relevant PIL (FPIL) \mmgg{constraint by} the location of flare pixels. Thereafter, planes at approximately equidistant locations along of the FPIL and oriented normally with respect to the local direction of the FPIL were defined, 
\rf{in order to derive the altitude of the \rff{current-weighted center ($\hcenter$) of the} MFR/SA.} Based on FPIL-normal planes and in addition ---as traditionally done--- based on an FPIL-aligned plane we estimated the critical height for torus instability ($\hc$) using the method described \rrff{in \href{ss:stability}{Sect.~\ref{ss:stability}}.} 

From the time series (\href{$hcmedian$_main}{Fig.~\ref{$hcmedian$_main}}), we \rf{computed time-}averaged \rf{pre-flare} values of \rf{$\mbf{\hcbar^{(n=1.5)}}$} 
over a 5-hour interval prior to the start of flares 
\rf{(last column in \href{t1}{Table~\ref{t1}}).}
In addition, we estimated time-averaged \rf{pre-flare} values of 
\rff{$\hcentermean$ (fourth column in \href{t1}{Table~\ref{t1}}).} 
\rf{
From \href{$hcmedian$_main}{Fig.~\ref{$hcmedian$_main}} (and the pre-flare time-averaged values in \href{t1}{Table~\ref{t1}}) it is evident that the structural characteristics of MFRs (SAs), as probed by our \rff{$\mbf{\hcentermean}$ estimates}, are not systematically different prior to confined versus eruptive flaring. From \href{$hcmedian$_main}{Fig.~\ref{$hcmedian$_main}}, however, it seems \rff{that $\mbf{\hcentermean}$ is} decreasing during the eruptive flares. This finding supports that the optimization-based modeling used here grasps the flare-related transformation of \rrff{MFRs} (SAs) to some extent.} 

\rf{
Noteworthy, our estimates of $\hcentermean$ are found restricted to low altitudes in the model volumes ($\approx$5\,Mm), far below the critical height for TI (\rf{$\gtrsim$13\,Mm} for $\nc$\,=\,1.5, the critical value most often used in literature). This would imply that none of the analyzed flares, including the obviously eruptive ones, were prone to TI. This being rather unlikely (and contrasting the conclusions drawn in a number of individual studies on the target ARs in our sample), hints at a different possibility. Using $\nc$\,=\,1.5 as a threshold as commonly done must be regarded as very restrictive, however, especially given the underlying assumption of a toroidally symmetric, large-aspect-ratio flux rope in \cite{2006PhRvL..96y5002K}, and given that any departure from such an idealized structure significantly effects the forces acting on a MFR \citep{2016PhPl...23k2102M}. Correspondingly, from the observation-based modeling survey of \cite{2019ApJ...884...73D}, the authors suggested $\nc$\,=\,1.3 as a more realistic threshold (with the possible impact of particular methods used to model the coronal magnetic field unknown to date). Ultimately, there might be a range of values at which coronal structures become unstable, down \rff{to $\nc$\,=\,0.4 and up to $\nc$\,$\sim$\,2.5 \citep[e.g.,][]{2019ApJ...884..157Z}.} In order to address this aspect, we inspect $\mbf{\hcentermean}$ with respect to $\mbf{\hcbar}^{(n=0.8)}$ \rff{\citep[based on][]{2016PhPl...23k2102M,2018ApJ...864..138J}} assuming to be representative for a lower corresponding limit. From \href{t1}{Table~\ref{t1}}, it appears that using this lower threshold value for $\mathbf{\nc}$ yields a distribution of $\mbf{\hcang^{(n=0.8)}}$ that also to some extent segregates confined from eruptive flares (mean of $\approx$\,32\,Mm and $\approx$\,8\,Mm, respectively, see third column in the Table) and that $\mbf{\langle\hcentermean\rangle}$ reaches on average to that height regime prior to eruptive flares ($\deltah$\,$\lesssim$\,\rff{4\,Mm}, on average, \rrff{fifth column of the table)}. \rrff{This gap ($\deltah$) between $\langle\hcentermean\rangle$ and $\mbf{\hcang}^{(n=0.8)}$ is exacerbated by the averaging done over the FPIL-normal planes which results in slightly lower values of the current-weighted center and higher $\mbf{\hcbar}^{(n=0.8)}$ than the $\hc^{(n=0.8)}$ values at the position of the current-weighted center of the non-potential structure (black cross in \href{bz_map}{Figs.~\ref{bz_map}}, \href{jmapa}{~\ref{jmapa}} and \href{jmapb}{~\ref{jmapb}})}}.

\rf{Following, \nrf{for example}, \cite{2015Natur.528..526M} and \cite{2019ApJ...871..105Z}, there might be even cases where torus-unstable MFRs just fail to \rff{erupt fully}. In order to possibly spot such cases, we interpret our findings also in context with the type of the confined flares --- involving a reconfiguration without significant eruption or a failed eruption confined at larger heights \citep[type I and II, respectively, see][]{2019ApJ...881..151L}). For confined events of type I (the X3.1 flare in AR 12192 and the X1.5 flare in AR 11166) for which TI can be excluded as a flare trigger, we indeed find $\mbf{\hcbar}>>\hcentermean$, irrespective of what threshold is used ($\nc=0.8$ or $\nc$\,=\,1.5; see  \href{$hcmedian$_main}{\rrff{Figs.}~\ref{$hcmedian$_main}}(b) and \href{$hcmedian$_main}{\ref{$hcmedian$_main}}(j), respectively), serving as examples of “true negatives”. For the confined events of type II (the analyzed M-class flares hosted by AR 12268, 11302, and 11339) also, $\mbf{\hcbar}>>\hcentermean$, as expected (see \href{$hcmedian$_main}{\rrff{Fig.}~\ref{$hcmedian$_main}(d), \href{$hcmedian$_main}{\ref{$hcmedian$_main}}(f), and \href{$hcmedian$_main}{\ref{$hcmedian$_main}}(h)}, respectively, and \href{t1}{Table~\ref{t1}}).
}

\rf{
Yet another possible implication of our finding of $\mbf{\hcentermean}$ being considerably lower than $\mbf{\hcbar^{(n=1.5)}}$ is a rather discouraging one, namely that the optimization-based NLFF modeling is not suited for TI-related analysis. In other words, the non-potential structures obtained based on optimization-based NLFF modeling succeed \rff{to picture the geometrical properties} of coronal MFR (SA) structures only partially. Indeed, the $\hcentermean$ values presented here are similar to the central axis height of the MFR recovered by optimization-based NLFF modeling of AR 12891 as examined in \cite{2023A&A...669A..72T}. In that study the coronal altitudes of the center/arcade height of the model MFR were compared to the stereoscopically reconstructed height of the simultaneously observed active-region filament/prominence. Noteworthy, the NLFF-based MFR center height was found to be located several Mm lower than the stereoscopic estimate of the filament/prominence height, while the estimated (apex) height of the confining magnetic arcade was found at a height similar to the stereoscopic one. Thus, the vertical extent of magnetic structures in optimization-based NLFF models should possibly be regarded as representing an under-estimation (lower limit) of the true one.}

\rf{
This is supported also by the comparison to the results based on alternative reconstruction techniques which may yield different model MFR configurations, including a different estimate of $\hcentermean$. For instance, the data-driven approach applied in \cite{2020ApJ...890...10Z} to model AR 12673, suggests a pre-X9-flare value of the model MFR center of $\approx$19\,Mm, 
with the authors taking the apex height of the magnetic field line with the strongest twist as a reference. From their Fig. 3(i) one might get the impression that this represents an extreme value (an upper limit), with the respective field line being located on the outskirts/envelope of the MFR body. Based on that figure, we hypothesize here that a corresponding twist-weighted estimate of the MFR center would possibly be found a few Mm lower (above $\approx$10\,Mm), yet still at a distinctly larger altitude than our $\hcentermean\approx5$\,Mm estimate based on our optimization-based NLFF model. Additional comparison to the set of events analyzed in \cite{2019ApJ...884...73D}, who used a data-driven scheme to model the coronal magnetic field above all major-flare productive ARs of solar cycle 24, we may safely assume that our optimization-based MFR structures are residing at model heights lower by a factor of roughly two \rrff{\citep[see right column of Figs. 3 -- 9 in][]{2019ApJ...884...73D}}.}

\rf{Based on the above considerations, we are now able to inspect the interplay of local and global stability measures.} The helicity ratio, being based on the volume-integrated magnetic helicity, represents a global measure for the eruptive potential of a solar AR (global meaning length scales typical for solar ARs). As discussed in \cite{2021A&A...653A..69G}, a helicity ratio exceeding a certain value (0.1) does not unambiguously hint at an eruptive flare to happen, \rf{but merely expresses that the considered AR posses the potential to produce one \citep[see also the recent work by][]{2023ApJ...945..102D}. To better understand this, also in relation to the more local impact of the strapping field, we interpret in the following the pre-flare \rrff{time-}averaged values of the helicity ratio and the critical height for TI.}

\begin{figure}[t]
    \captionsetup{width=\linewidth}
   \begin{subfigure}[t]{0.5\textwidth}
        \centering
        \includegraphics[width=1.0\linewidth]{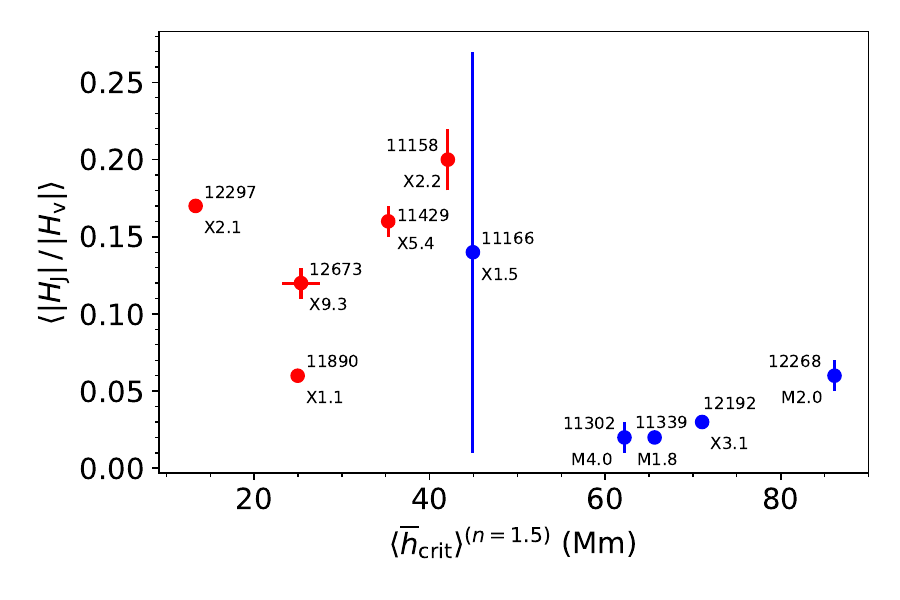}\\
    \end{subfigure}
    \caption{Scatter diagram of 
    \rf{pre-flare averaged helicity ratio, \rf{$\mhjprime$} \mmgg{\citep[see Table~3 of][]{2021A&A...653A..69G}}}
    versus 
    \rrff{pre-flare time-averaged critical height, $\mbf{\hcang}$}, for $\nc$\,=\,1.5 as estimated from the FPIL-normal planes. Red (blue) bullets show eruptive (confined) flares, respectively.}
    \label{scatter2}
\end{figure}
 
\mmgg{Using the} \rf{pre-flare averaged helicity ratio ($\mhjprime$)} 
from Table 3 of \cite{2021A&A...653A..69G}, we generate a scatter plot 
\rf{with respect to the pre-flare averaged} critical height \rf{for TI (using $\nc=1.5)$} 
(\href{scatter2}{\rrff{Fig.}~\ref{scatter2}}). \rff{Both of the quantities were averaged over a time window of five hours prior to flare onset.} Based on critical heights, we find a segregation among the ARs that produced confined \mgg{versus} eruptive flares. The ARs that produced large confined flares, we find the average \rf{pre-flare} critical height above \rff{44 Mm}, and for the ARs that produced large eruptive flares it is below \rff{43 Mm} 
\rff{(see last column} of \href{t1}{Table.~\ref{t1}}, and \href{scatter2}{\rrff{Fig.}~\ref{scatter2}}), in line with earlier corresponding studies \citep[e.g.,][]{2017ApJ...843L...9W,2018ApJ...853..105B}. \rff{Also,} in the more recent study by
\cite{2022A&A...665A..37J} \rf{critical heights were most commonly observed in the range 40--50\,Mm prior to CME onset times}. In contrast to \rf{their study, however,} 
we do not find any characteristic trends in time profiles of the critical height prior to eruptive or confined flares, likely because of the short duration of the time series analyzed here. 

\rff{As discussed in \cite{2018ApJ...853..105B}, eruptive flares originate either from the periphery of ARs where the overlying field is generally weaker, or originate from within the strong confinement of a dipole field in compact ARs (above which the confining magnetic field decays rapidly with height). Four of the five eruptive events studied in our restricted sample fall into the latter category. Correspondingly, $\mbf{\hcang^{(n=1.5)}}$ and $\mbf{\hcang^{(n=0.8)}}$ are also found at lower coronal heights for eruptive events.} \rf{Furthermore, \href{scatter2}{\rrff{Fig.}~\ref{scatter2}} indicates that $\mbf{\hcang^{(n=1.5)}}$ is more (clearly) segregating in terms of flare type than is the helicity ratio. Moreover, two distinct groups appear:} one group with [$\langle\hjprime\rangle$\,$\gtrsim$\,0.1, $\mbf{\hcang^{(n=1.5)}}$\,$\lesssim$\,42\,Mm], 
characteristic for a \rrff{corona prior to eruptive flares}; and another group with [$\langle\hjprime\rangle$\,$\lesssim$\,0.1, $\mbf{\hcang^{(n=1.5)}}$\,$\gtrsim$\,42\,Mm], 
characteristic for the corona prior to confined flares. \rrff{This trend is also observed for the pre-flare time-averaged critical height obtained by using $\nc=0.8$ (i.e., $\mbf{\hcang}^{(n=0.8)}$; see third column of \href{t1}{Table.~\ref{t1}}).} \rff{We note that the large mean value and uncertainty of $\hjprime$ in case of AR 11166 is due to a relative helicity reversal from positive to marginally negative during a very-short period in the time series \citep[see \rrff{Figs.} 2(j) and 6(j) of][]{2021A&A...653A..69G}, resulting in a sharp increase in $\hjprime$ at that instance, while $\hjprime$ was found to be less than 0.1 during rest of preflare interval. So altogether} it appears that a coronal field configuration with a higher eruptive potential, as sensed by $\hjprime$, is also a configuration more prone to TI (as sensed by $\hcbar$). 

\section{Summary and conclusions}
We aimed to better understand the role of \rrff{the strapping} field in the occurrence of confined \rf{versus} eruptive \mgg{large} flares by the combined analysis of the \rff{geometry} of the underlying magnetic structure (a magnetic flux rope or sheared arcade, depending on the AR considered) and the critical height for TI. To do so, we analyzed the model magnetic field and electric current density projected into several vertical planes, either aligned with or normal to the flare-relevant PIL, based on NLFF magnetic field modeling of ten active regions. An important new aspect in our analysis is that we address the interplay of local (PIL-associated as sensed by $\hc$) and global (active-region) eruptive potential of the ARs, the latter characterized by the helicity ratio.

\rrff{Similar} to earlier dedicated works, we find a clear segregation between confined and eruptive flares in terms of $\hc$ for TI, rather than regarding a characteristically different pre-flare magnetic field (in terms of \rff{the altitude of the current-weighted center} of the pre-flare core field). In particular, we find the torus-unstable regime to reside lower down in the model atmosphere, closer to the altitudes occupied by the non-potential core field prior to eruptive flares. Importantly, in general it appears that magnetic field configurations which adhere to a generally larger eruptive potential (as sensed by the helicity ratio) are more prone to TI.

This most important finding awaits confirmation from the examination of a larger sample of events, and also in respect to the investigation of effects complicating the analysis. This includes the impact of, for instance, the selected approach to approximate the flare-relevant PIL and corona above of it, the method used to detect geometry-related measures (e.g., altitude of central magnetic axis or strongest electric currents, etc.) of the non-potential core field (a flux rope or sheared arcade), and also clarification of whether the coronal magnetic field modeling used (optimization-based NLFF modeling in this study) qualifies for such an analysis. The latter is possibly to be accomplished only by case-to-case verification of the model-deduced core magnetic field geometry by, \nrf{for example}, stereoscopic means.

\begin{acknowledgements} 
\rf{We thank the anonymous referee for critical yet valuable suggestions to improve our manuscript.}
This research was funded in part by the Austrian Science Fund (FWF) 10.55776/P31413. For the purpose of open access, the author has applied a CC BY public copyright license to any Author Accepted Manuscript version arising from this submission. SDO data are courtesy of the NASA/SDO AIA and HMI science teams.
\end{acknowledgements}

\bibliographystyle{aa} 
\bibliography{main} 

\begin{appendix}
\section{Additional Figures}

\begin{figure*}[htp]
    \captionsetup{width=\linewidth}
    \vspace{1.1cm}
    \begin{subfigure}[ht]{0.40\textwidth}
        \centering
        \begin{picture}(100,100)
        \put(0,0){\includegraphics[width=0.95\linewidth,trim={0cm 1.5cm 12cm 3cm},clip]{plots_new//magnetograms//noaa_12673_norm_width_full_58.32Mm_PIL__flare_ribbon06_0859.pdf}}
        \put(34,110){a) AR 12673}
        \end{picture}
    \end{subfigure}\hspace{1em}
    \begin{subfigure}[ht]{0.4\textwidth}
        \centering
        \begin{picture}(100,100)
        \put(0,0){\includegraphics[width=1\linewidth,trim={0cm 0cm 8cm 3cm},clip]{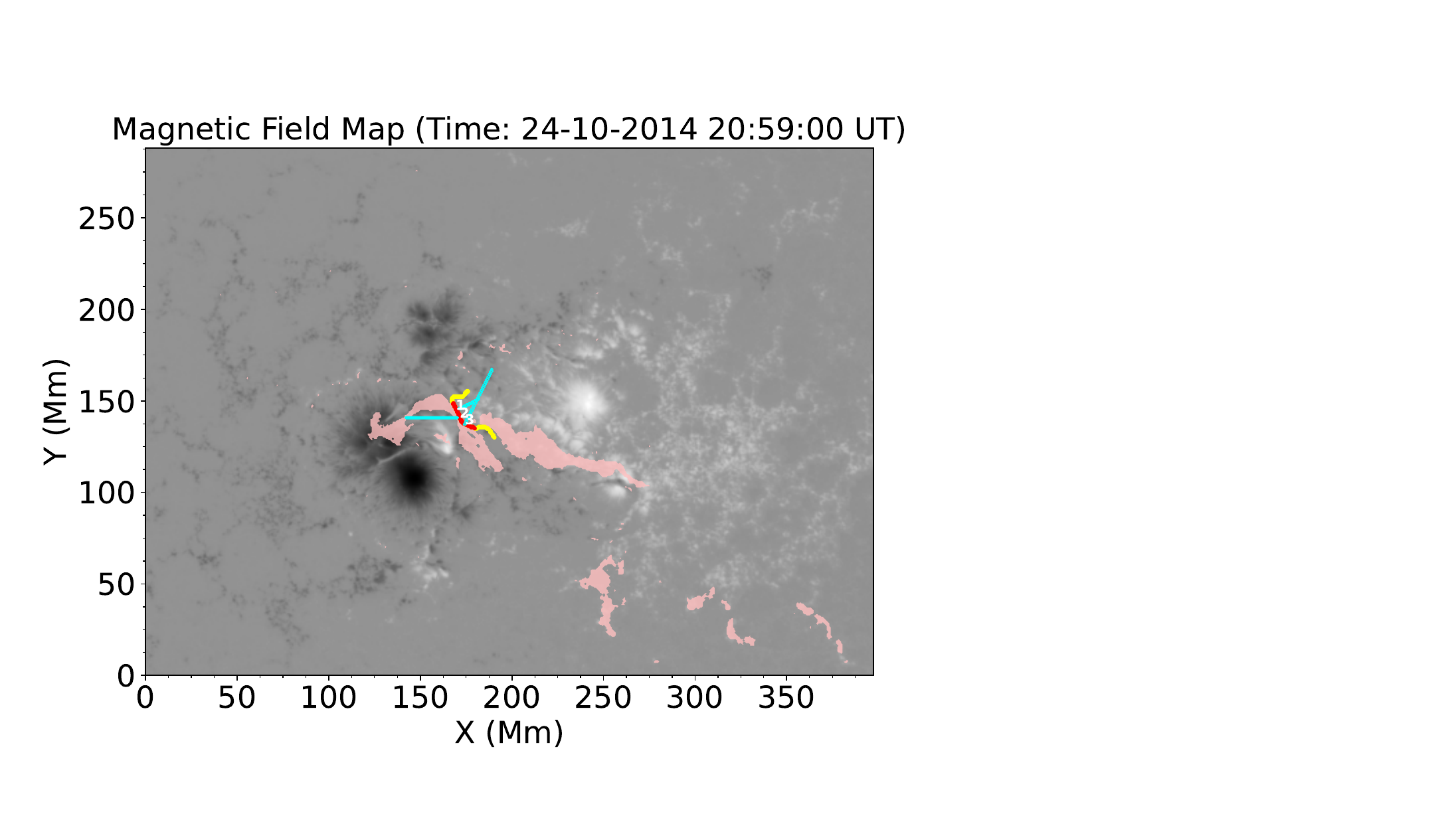}}
        \put(34,108){b) AR 12192}
        \end{picture}
    \end{subfigure}
    
    \vspace{1.2cm}\begin{subfigure}[ht]{0.4\textwidth}
        \centering
        \begin{picture}(100,100)
        \put(0,0){\includegraphics[width=0.95\linewidth,trim={0cm 0cm 12cm 3cm},clip]{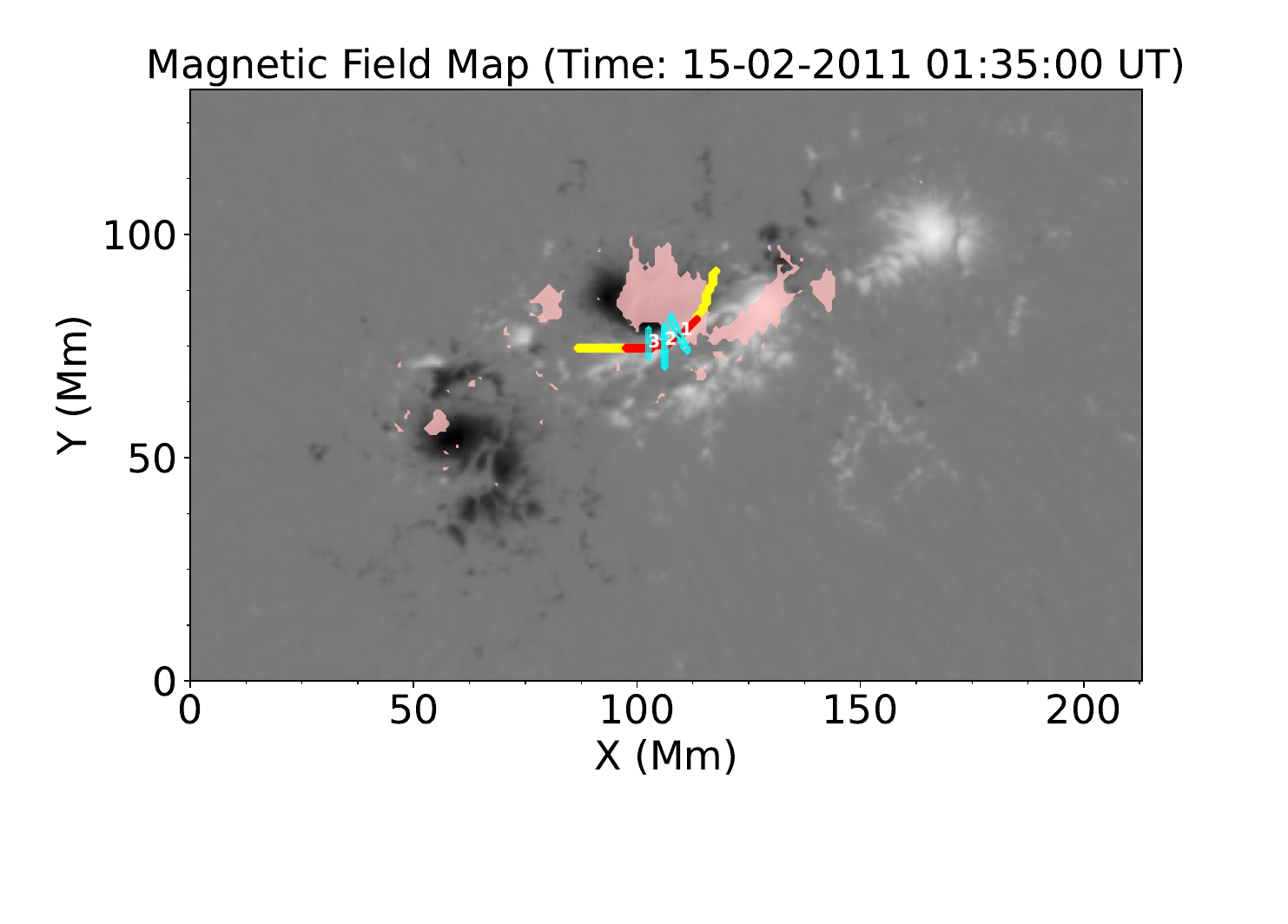}}
        \put(34,120){c) AR 11158}
        \end{picture}
    \end{subfigure}\hspace{1em}
    \begin{subfigure}[ht]{0.4\textwidth}
        \centering
        \begin{picture}(100,100)
        \put(0,0){\includegraphics[width=1.0\linewidth,trim={0cm 0cm 12cm 3cm},clip]{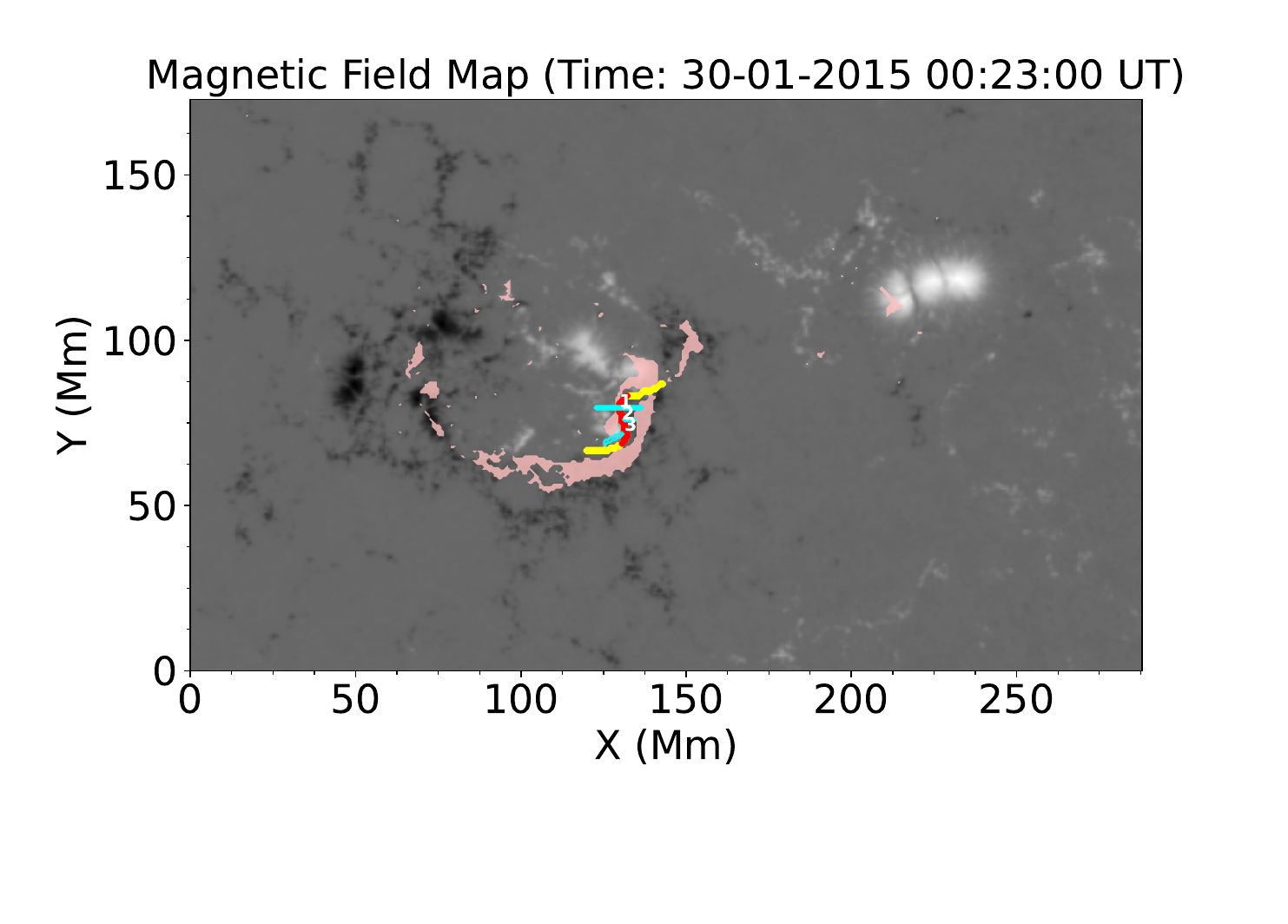}}
        \put(34,120){d) AR 12268}
        \end{picture}
    \end{subfigure}
    
    \vspace{0.8cm}\begin{subfigure}[ht]{0.4\textwidth}
        \centering
        \begin{picture}(100,100)
        \put(0,0){\includegraphics[width=0.95\linewidth,trim={0cm 0cm 12cm 3cm},clip]{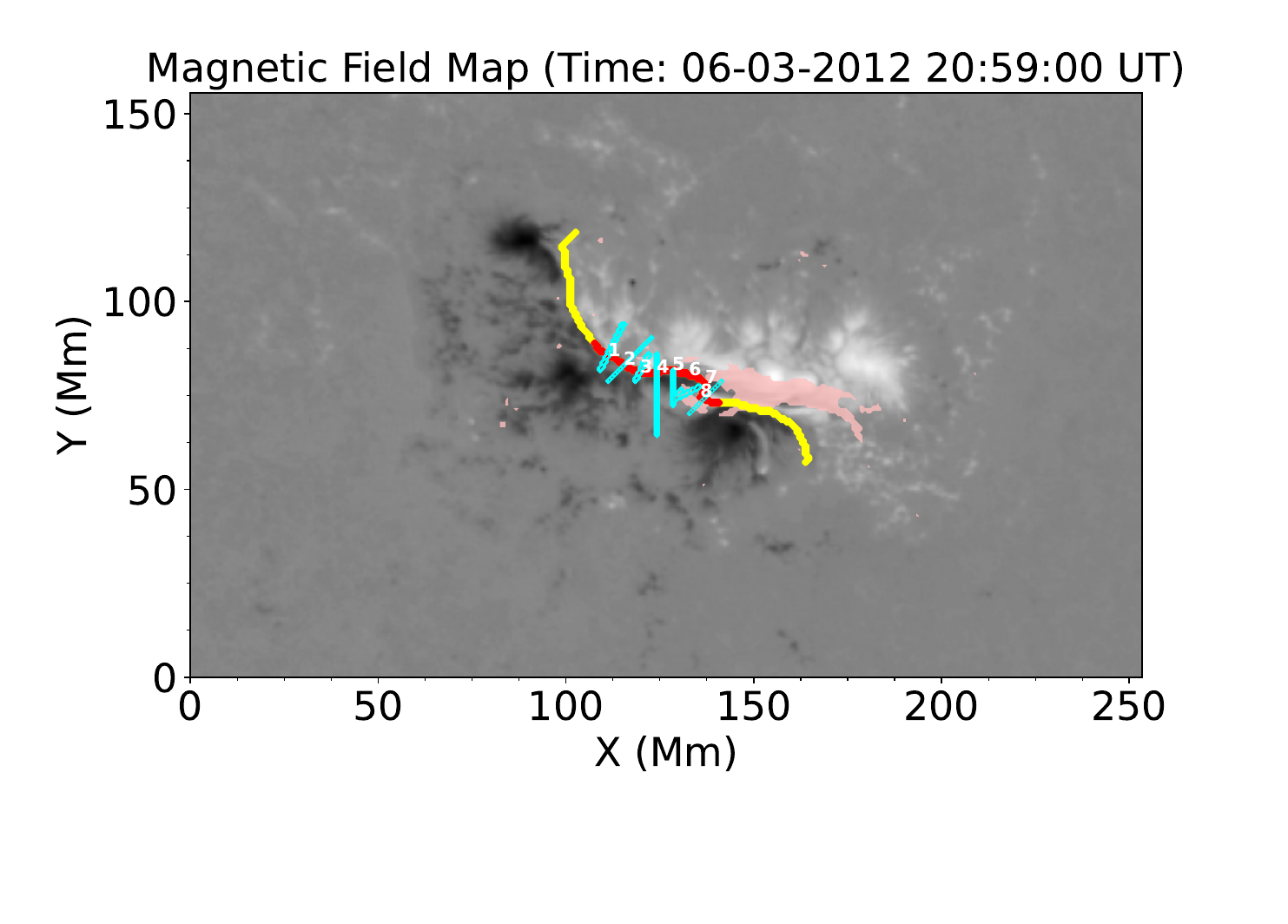}}
        \put(34,120){e) AR 11429}
        \end{picture}
    \end{subfigure}\hspace{1em}
    \begin{subfigure}[ht]{0.4\textwidth}
        \centering
        \begin{picture}(100,100)
        \put(0,0){\includegraphics[width=1.0\linewidth,trim={0cm 0cm 12cm 3cm},clip]{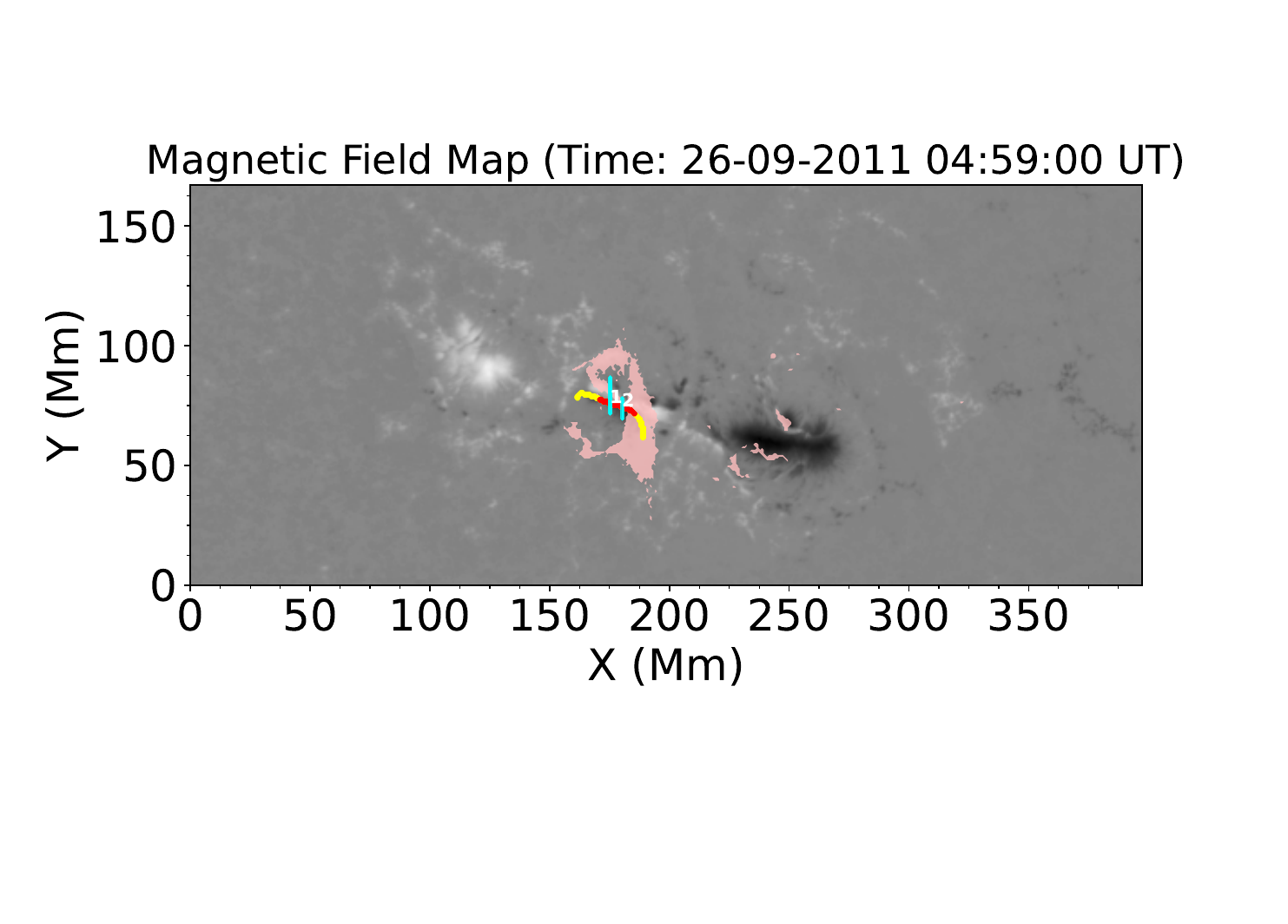}}
        \put(34,110){f) AR 11302}
        \end{picture}
    \end{subfigure}
    
    \vspace{1cm}\begin{subfigure}[ht]{0.4\textwidth}
        \centering
        \begin{picture}(100,100)
        \put(0,0){\includegraphics[width=0.95\linewidth,trim={0cm 0cm 12cm 2cm},clip]{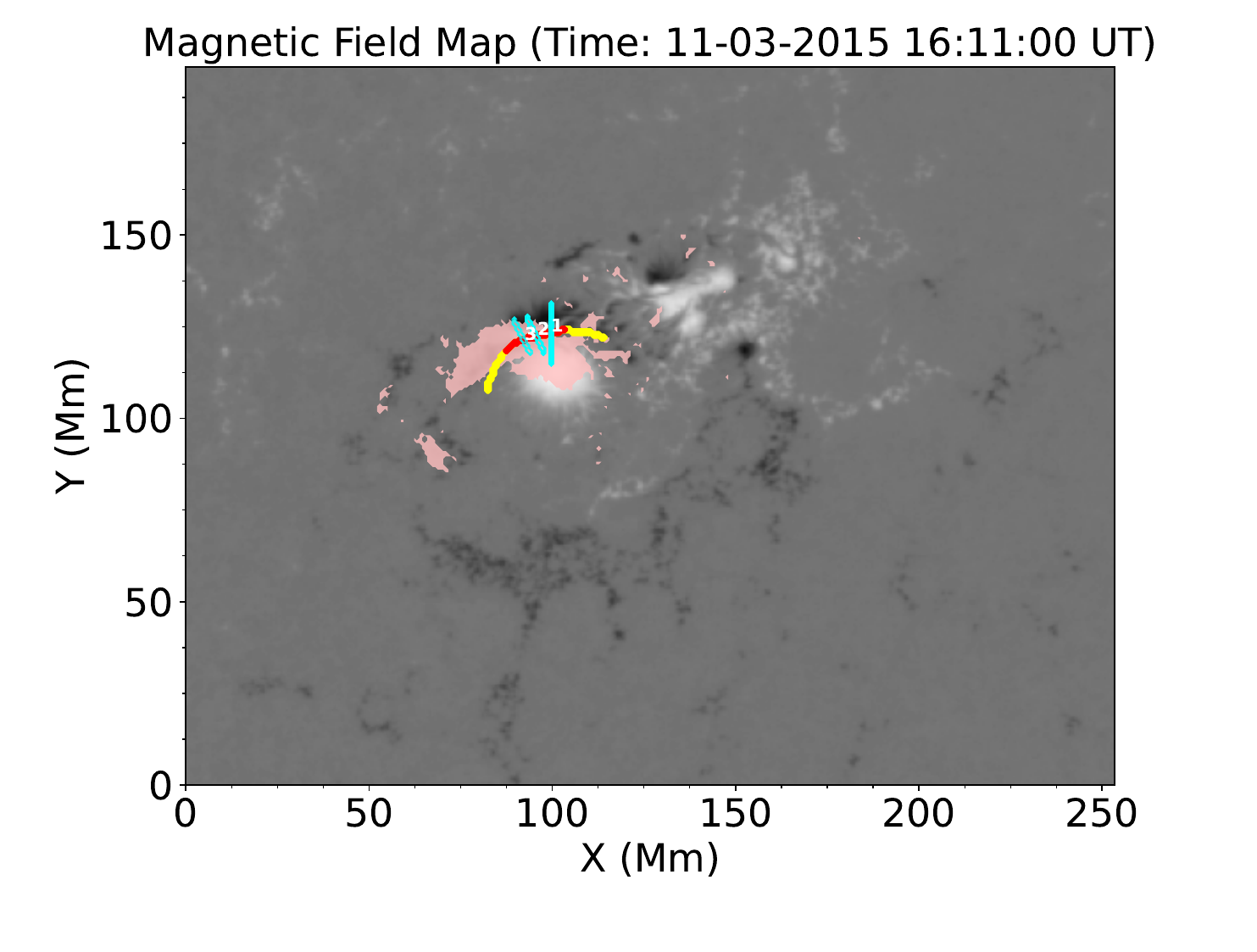}}
        \put(34,132){g) AR 12297}
        \end{picture}
    \end{subfigure}\hspace{1em}
    \begin{subfigure}[ht]{0.4\textwidth}
        \centering
        \begin{picture}(100,100)
        \put(0,0){\includegraphics[width=1.0\linewidth,trim={0cm 0cm 12cm 3cm},clip]{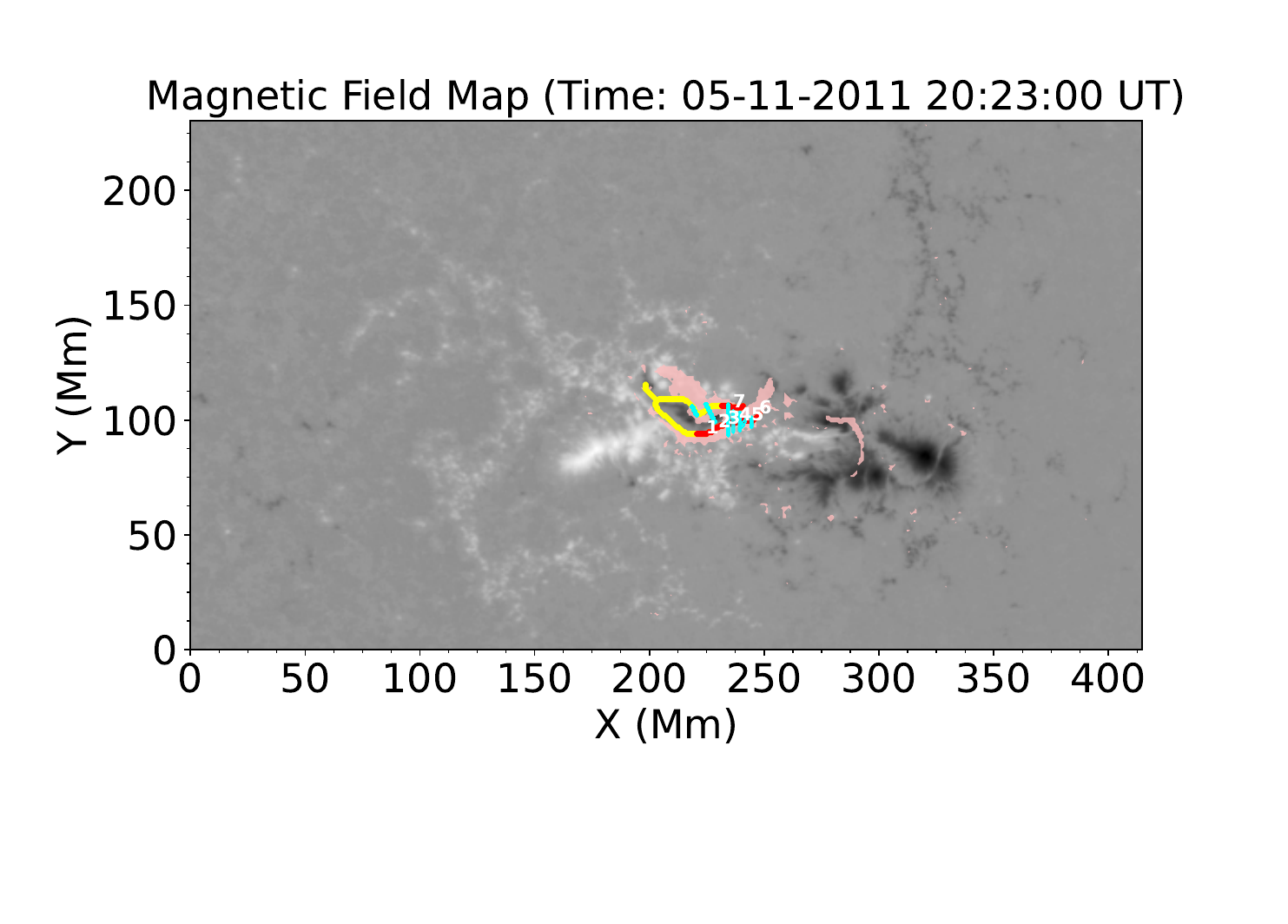}}
        \put(34,120){h) AR 11339}
        \end{picture}
    \end{subfigure}
    
    \vspace{1.3cm}\begin{subfigure}[ht]{0.4\textwidth}
        \centering
        \begin{picture}(100,100)
        \put(0,0){\includegraphics[width=0.95\linewidth,trim={0cm 0cm 12cm 2cm},clip]{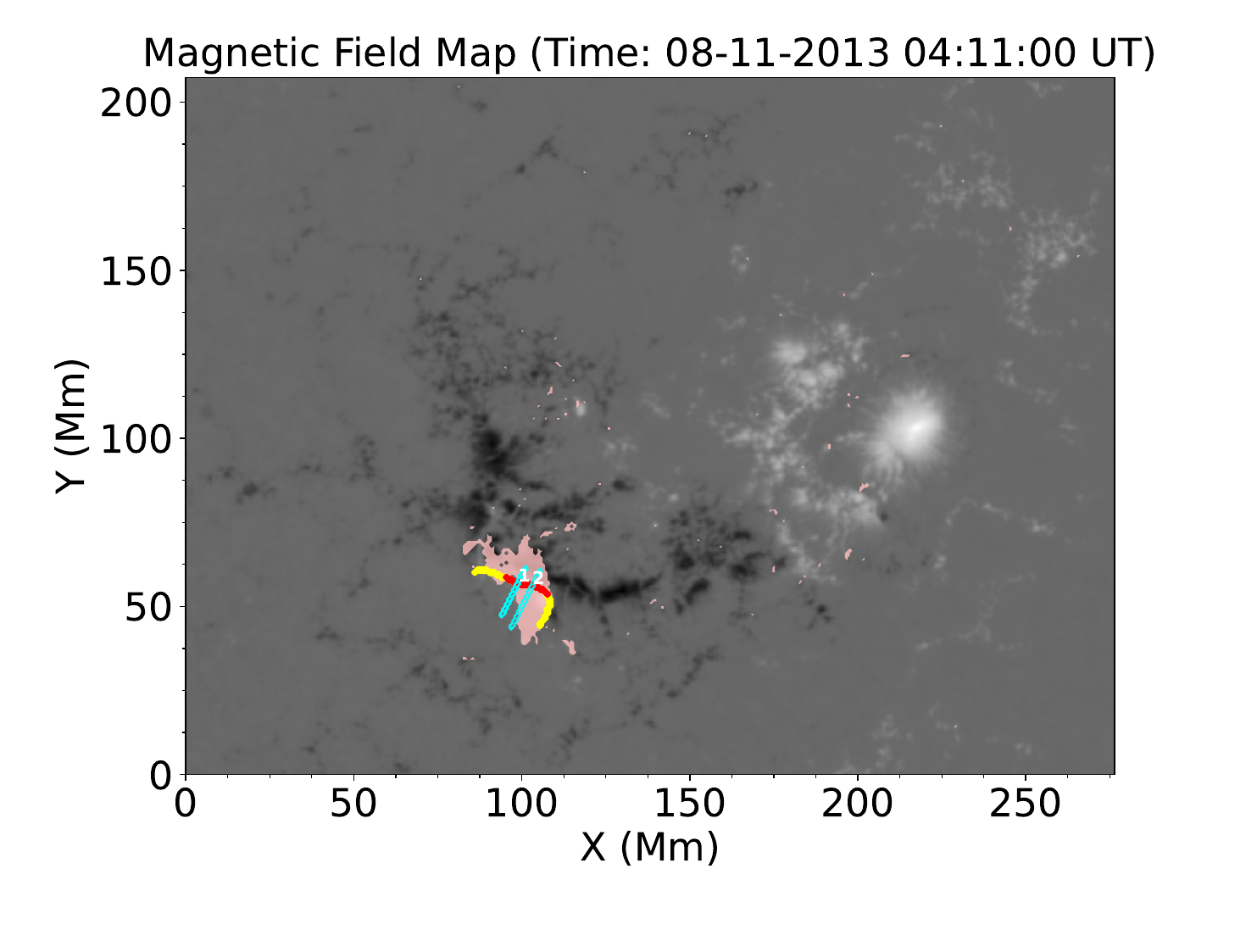}}
        \put(34,128){i) AR 11890}
        \end{picture}
    \end{subfigure}\hspace{1em}
    \begin{subfigure}[ht]{0.4\textwidth}
        \centering
        \begin{picture}(100,100)
        \put(0,0){\includegraphics[width=1.0\linewidth,trim={0cm 0cm 12cm 3cm},clip]{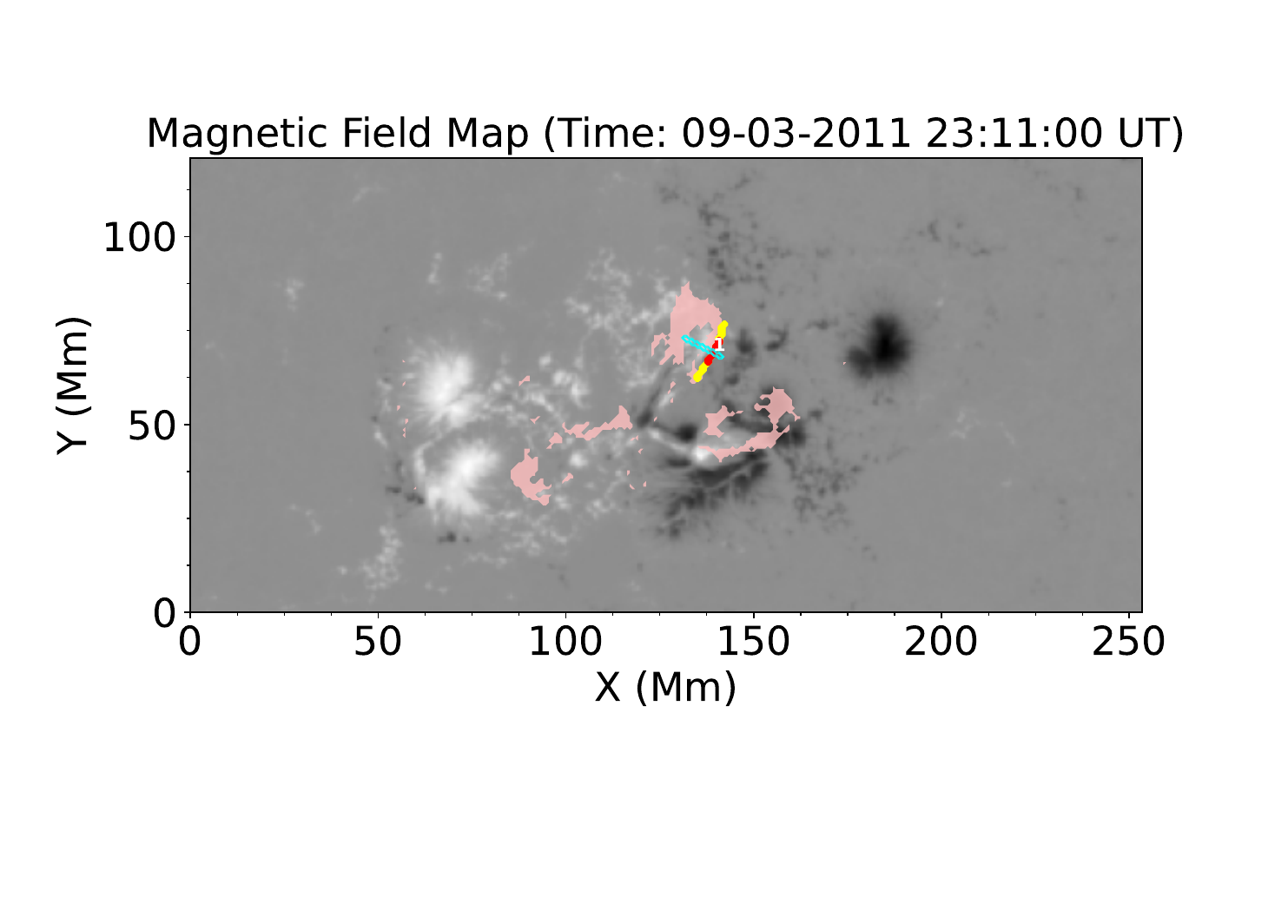}}
        \put(34,110){j) AR 11166}
        \end{picture}
    \end{subfigure}
    
    \caption{Photospheric magnetic field ($\Bz$) of the 10 ARs with flare ribbon (pink), flare-relevant (red) PIL along of the full detected PIL (yellow curve) and FPIL-normal planes (cyan lines) displayed on top of the magnetic field.}
    \label{app1}
\end{figure*}
\begin{figure*}[htp]
    \captionsetup{width=\linewidth}
    \hspace{1cm}
    \begin{subfigure}[ht]{0.45\textwidth}
        \centering
        \includegraphics[width=0.7\linewidth,trim={0cm 0cm 2cm 1cm},clip]{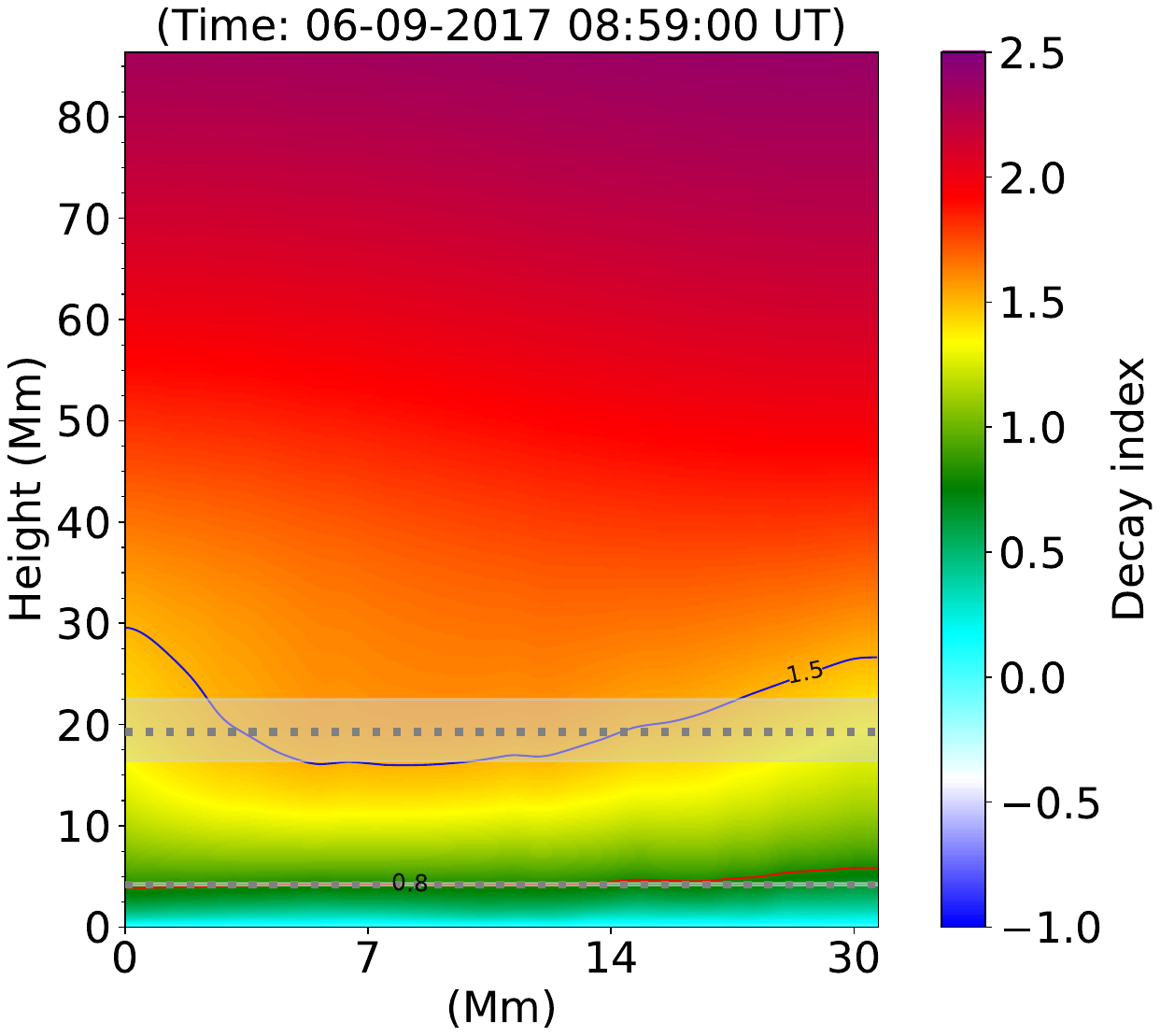}
        \put(-200,100){a) AR 12673}
    \end{subfigure}
    \begin{subfigure}[ht]{0.45\textwidth}
        \centering
        \includegraphics[width=0.8\linewidth,trim={0cm 0cm 0cm 0cm},clip]{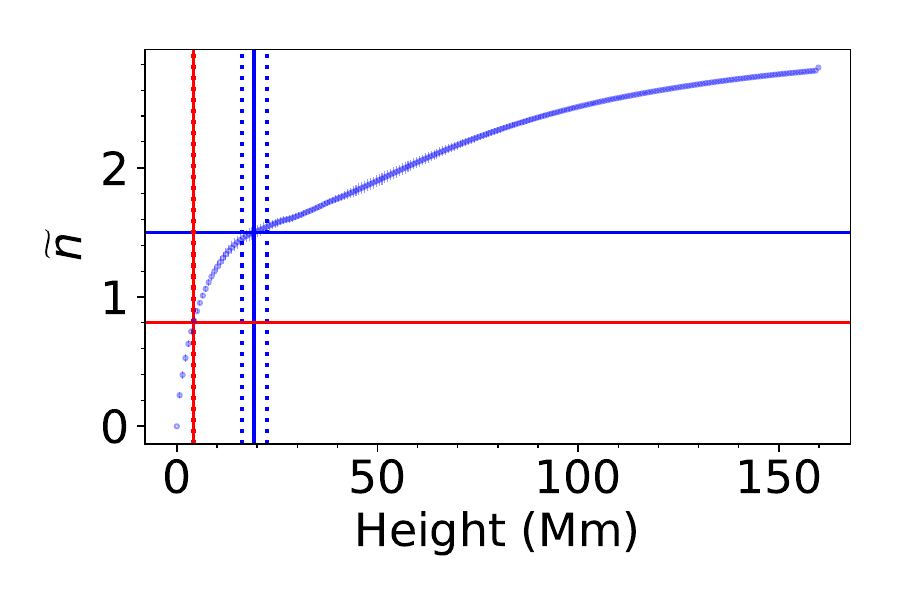}
        \put(-200,100){b)}
    \end{subfigure}
    
    \vspace{-0.3cm}\hspace{1cm}\begin{subfigure}[ht]{0.45\textwidth}
        \centering
        \includegraphics[width=0.7\linewidth,trim={0cm 0cm 2cm 1cm},clip]{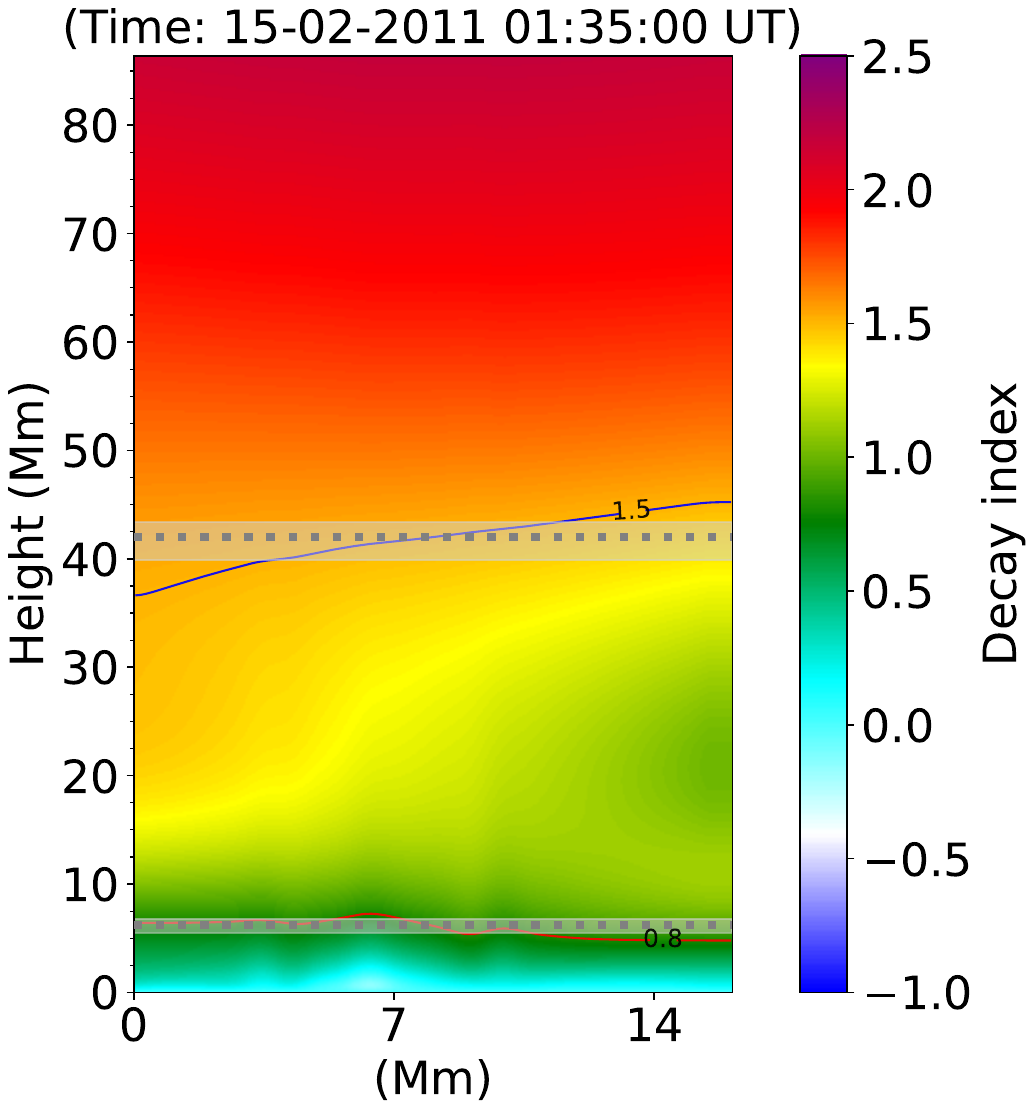}
        \put(-200,100){c) AR 11158}
    \end{subfigure}
    \begin{subfigure}[ht]{0.45\textwidth}
        \centering
        \includegraphics[width=0.8\linewidth,trim={0cm 0cm 0cm 0cm},clip]{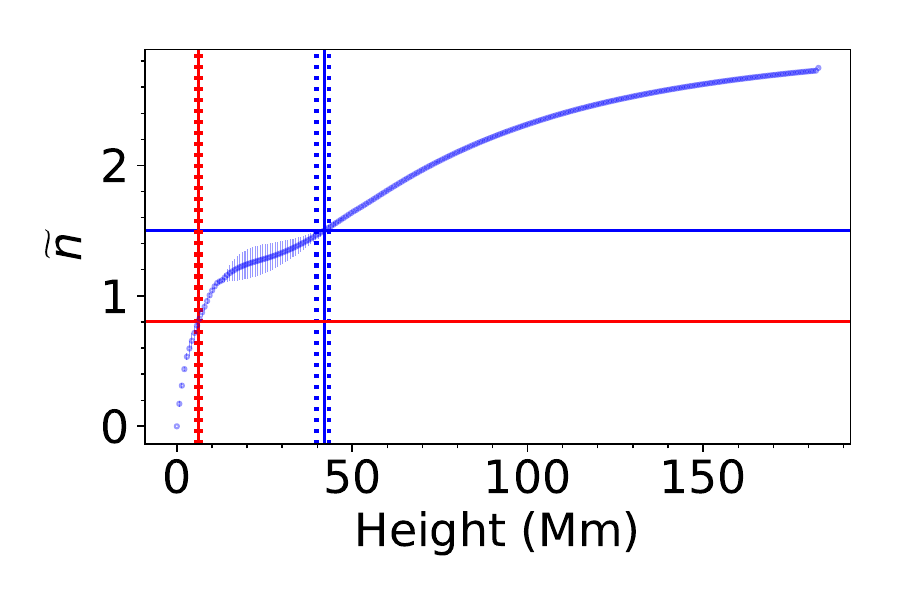}
        \put(-200,100){d)}
    \end{subfigure}
    
    \vspace{-0.35cm}\hspace{1.0cm}\begin{subfigure}[ht]{0.45\textwidth}
        \centering
        \includegraphics[width=0.7\linewidth,trim={0cm 2cm 2cm 2cm},clip]{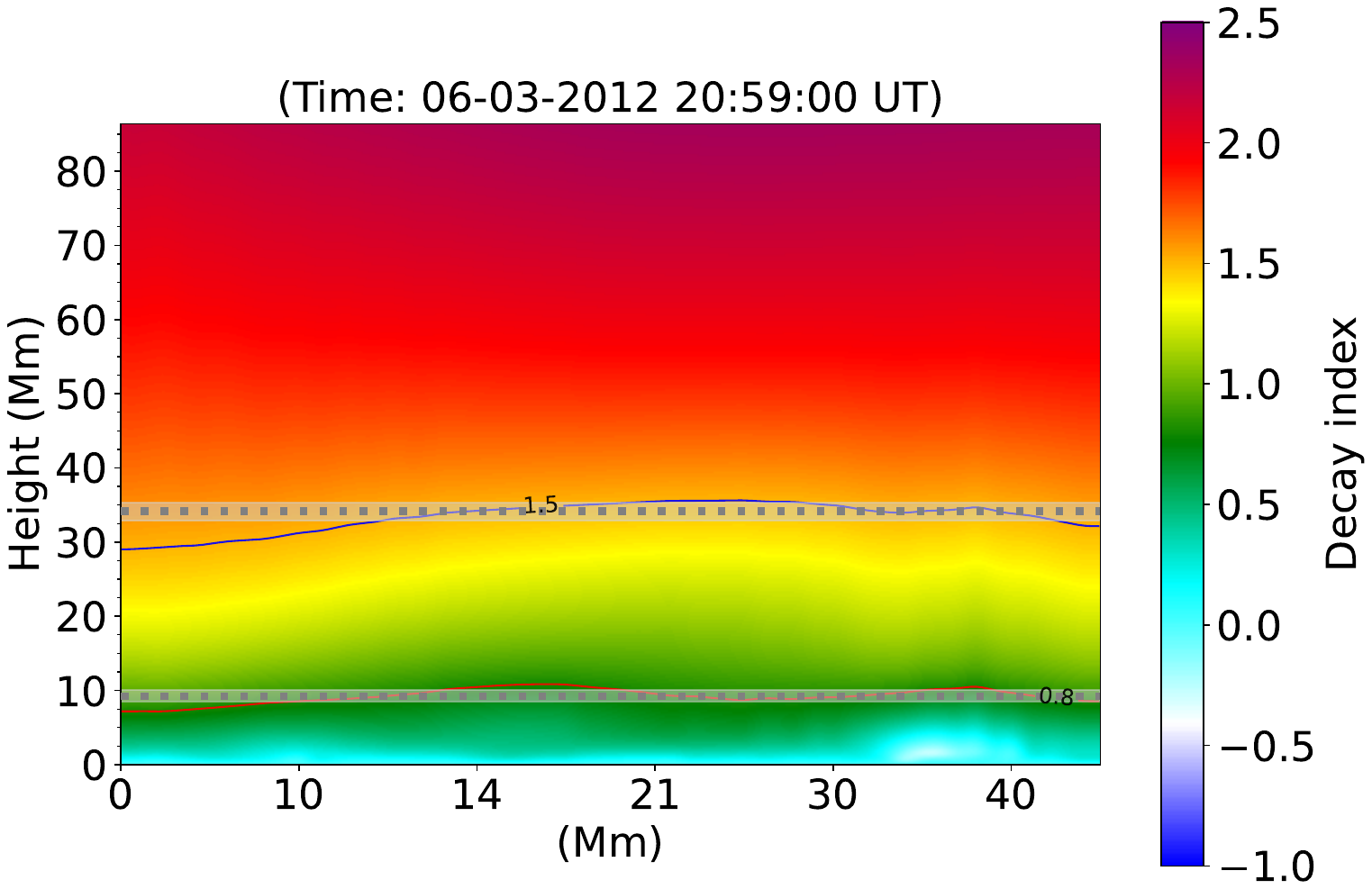}
        \put(-200,100){e) AR 11429}
    \end{subfigure}
    \begin{subfigure}[ht]{0.45\textwidth}
        \centering
        \includegraphics[width=0.8\linewidth,trim={0cm 0cm 0cm 0cm},clip]{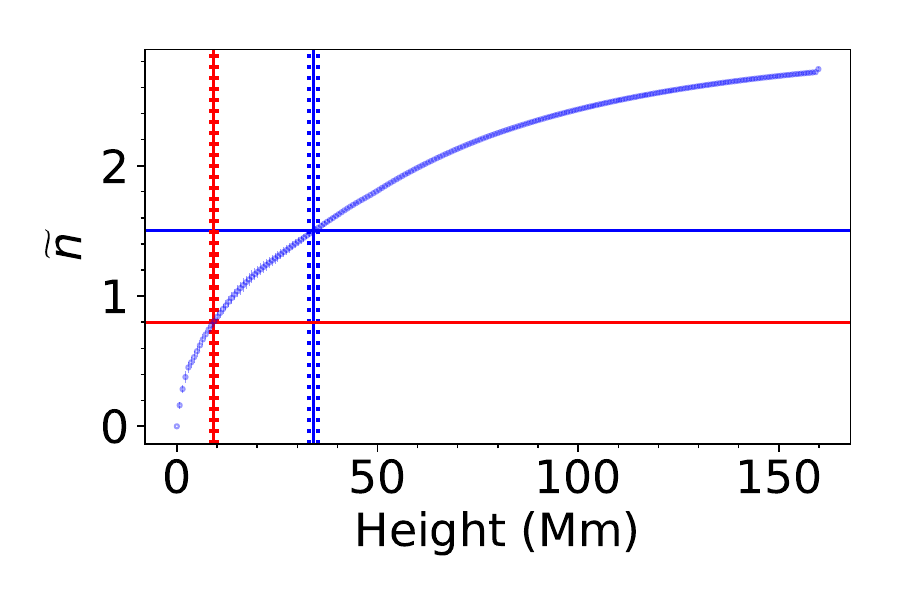}
        \put(-200,100){f)}
    \end{subfigure}
    
    \vspace{-0.35cm}\hspace{1cm}\begin{subfigure}[ht]{0.45\textwidth}
        \centering
        \includegraphics[width=0.7\linewidth,trim={0cm 0cm 2cm 1cm},clip]{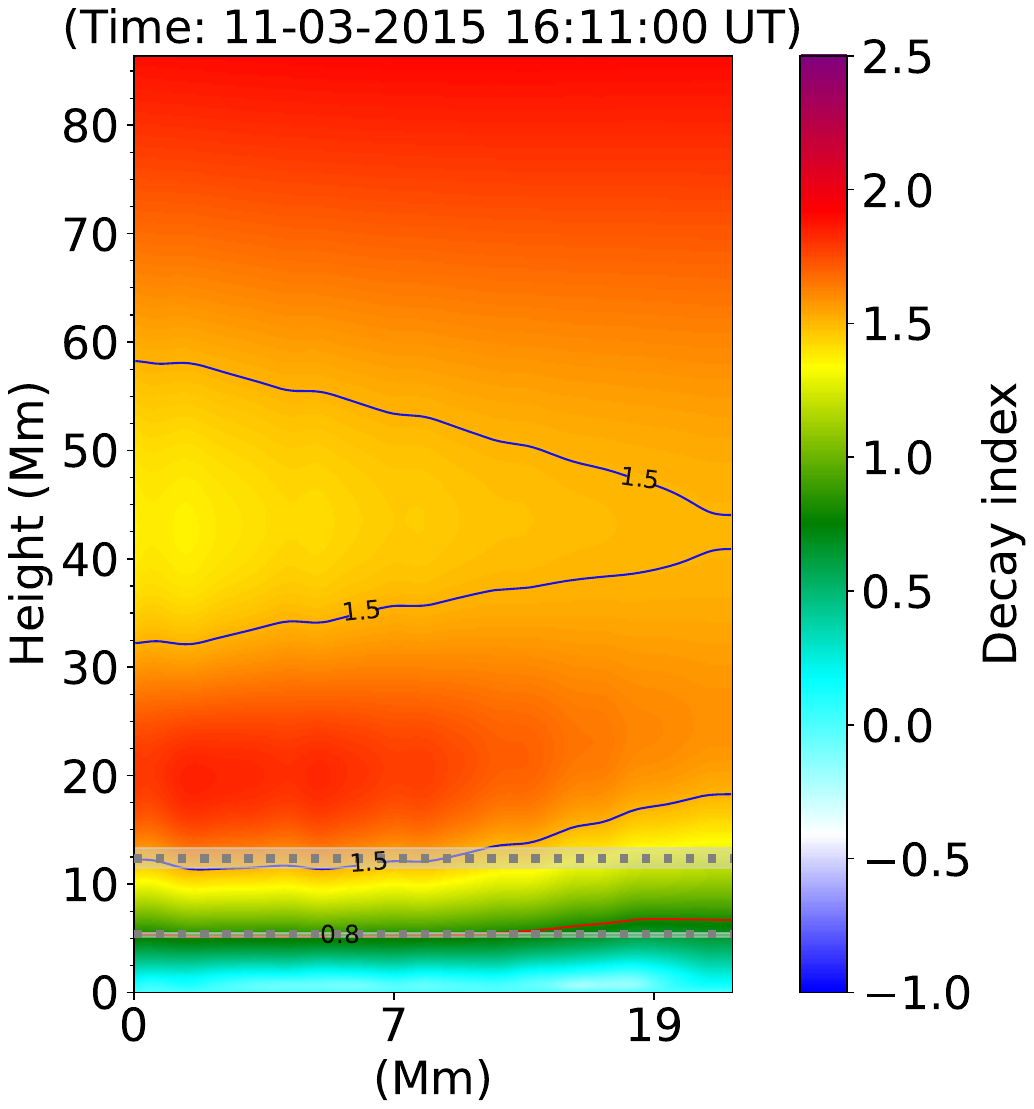}
        \put(-200,100){g) AR 12297}
    \end{subfigure}
    \begin{subfigure}[ht]{0.45\textwidth}
        \centering
        \includegraphics[width=0.8\linewidth,trim={0cm 0cm 0cm 0cm},clip]{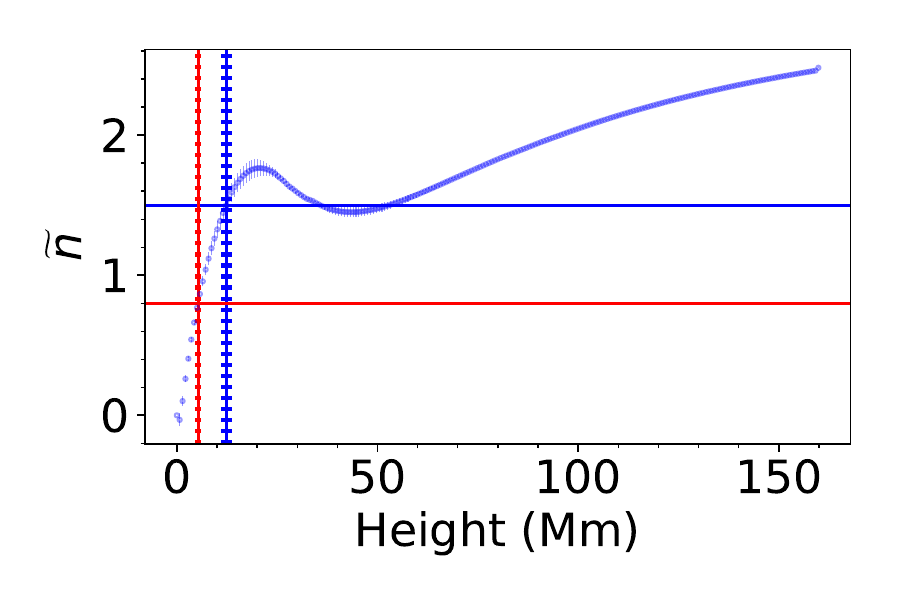}
        \put(-200,100){h)}
    \end{subfigure}
    
    \vspace{-0.25cm}\hspace{1cm}\begin{subfigure}[ht]{0.45\textwidth}
        \centering
        \includegraphics[width=0.8\linewidth,trim={0cm 0cm 0cm 1cm},clip]{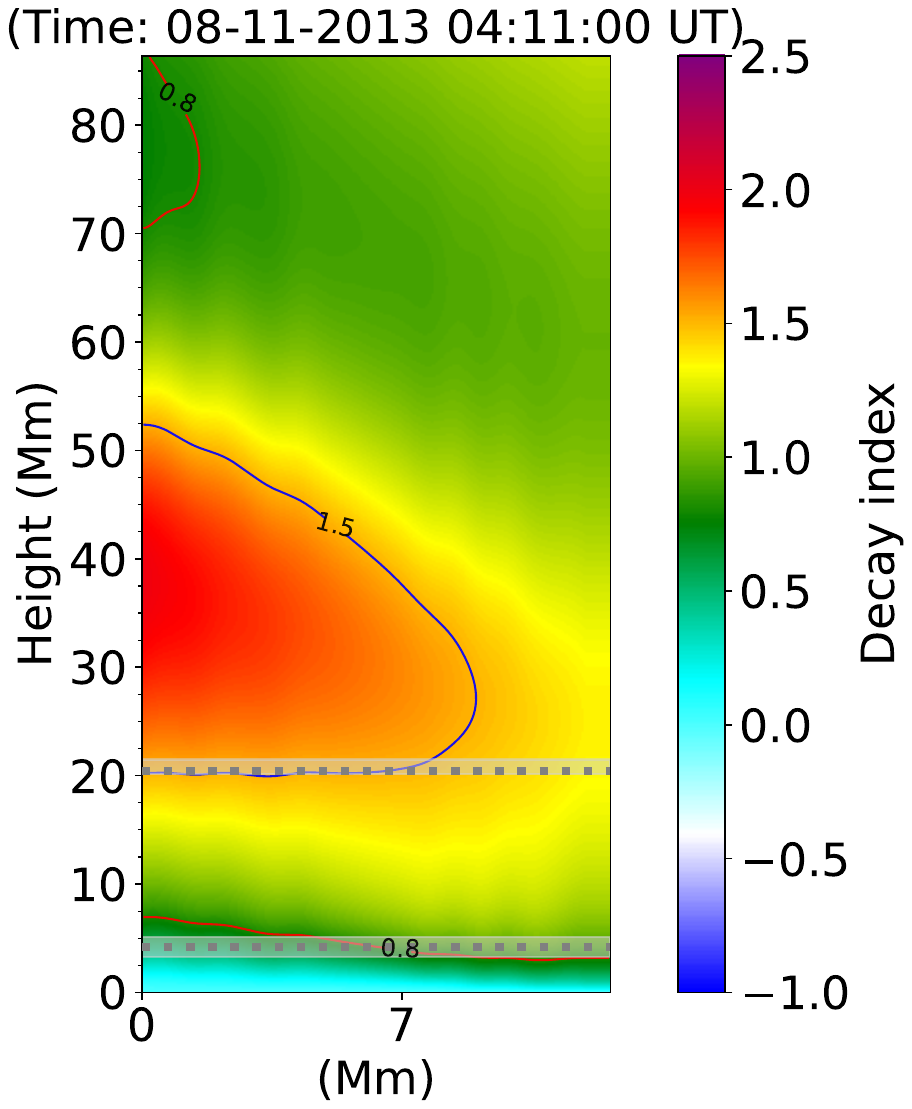}
        \put(-210,100){i) AR 11890}
    \end{subfigure}
    \begin{subfigure}[ht]{0.45\textwidth}
        \centering
        \includegraphics[width=0.8\linewidth,trim={0cm 0cm 0cm 0cm},clip]{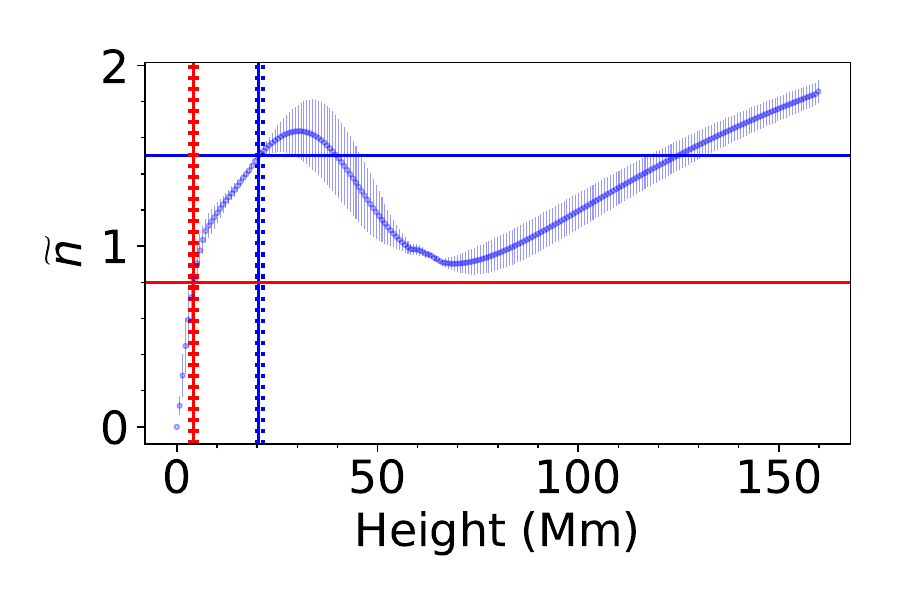}
        \put(-200,100){j)}
    \end{subfigure}
    \caption{
    \rf{{\it Left column:} Distribution of the decay index in the FPIL-aligned planes.} The blue and \rf{red} curves indicate \rf{where $n$ exceeds 1.5 and 0.8, respectively. Gray dotted lines indicate the correspondingly estimated critical height for TI. Shaded areas mark the corresponding uncertainties.}
    {\it Right column:} Profile of $\mbf{\nmedian}$ vs.\ height. The blue (red) horizontal lines indicate where \rff{$\mbf{\nmedian}$\,=\,1.5 (0.8)}, respectively. Vertical blue solid (dotted) lines mark the heights at which $\mbf{\nmedian}$\,$\pm$\,1\,MAD=1.5. Vertical red solid (dotted) lines mark the heights at which $\mbf{\nmedian}$\,$\pm$\,1\,MAD=0.8.} 
    \label{app1_1}
\end{figure*}

\begin{figure*}[htp]
    \captionsetup{width=\linewidth}
    \hspace{0.75cm}\begin{subfigure}[ht]{0.45\textwidth}
        \centering
        \includegraphics[width=0.8\linewidth,trim={0cm 0cm 0cm 0cm},clip]{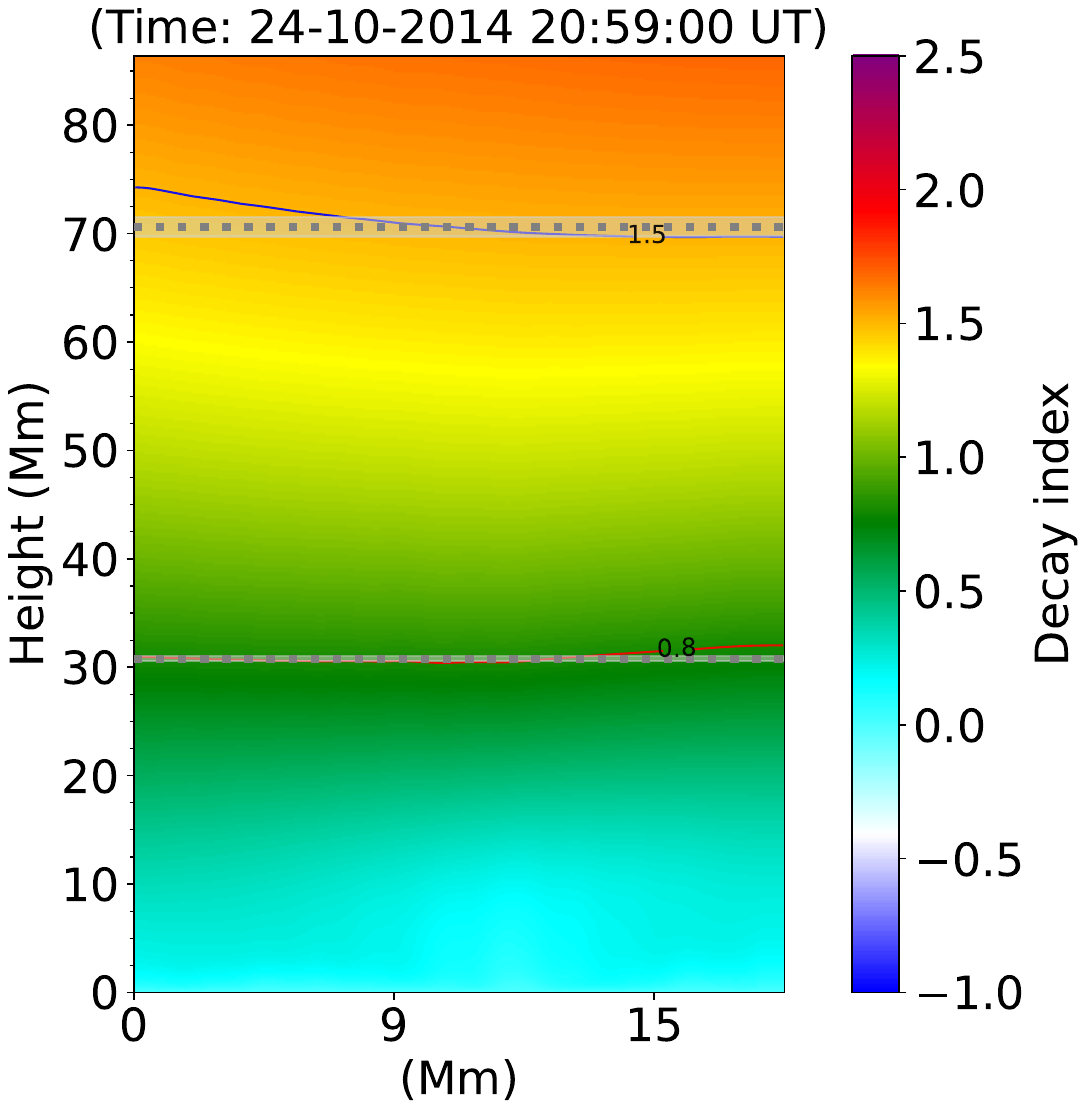}
        \put(-185,107){a) AR 12192}
    \end{subfigure}\hspace{-1em} 
    \begin{subfigure}[ht]{0.45\textwidth}
        \centering
        \includegraphics[width=0.8\linewidth,trim={0cm 0cm 0cm 0cm},clip]{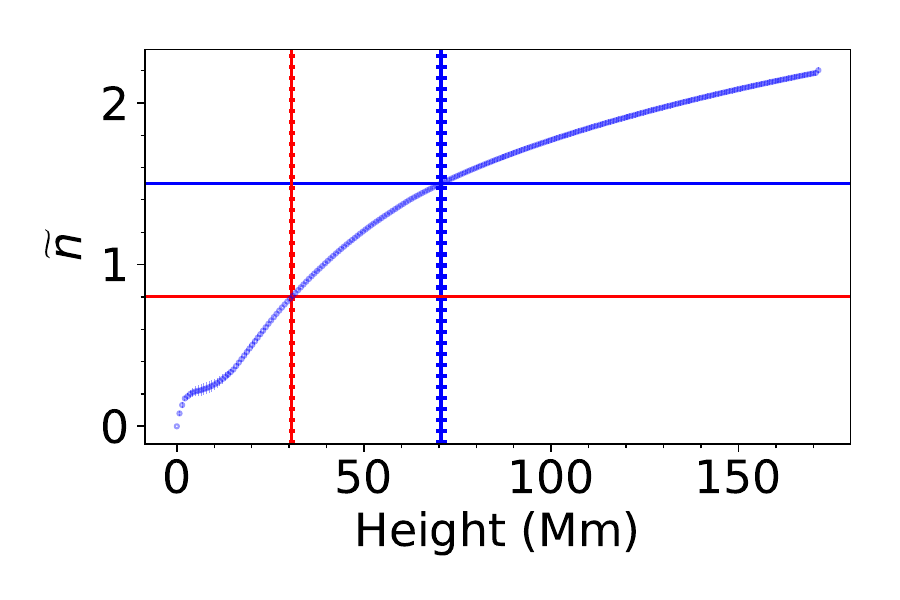}
        \put(-200,100){b)}
    \end{subfigure}

    \vspace{-0.2cm}\hspace{0.5cm}\begin{subfigure}[ht]{0.45\textwidth}
        \centering
        \includegraphics[width=0.9\linewidth,trim={0cm 0cm 0cm 0cm},clip]{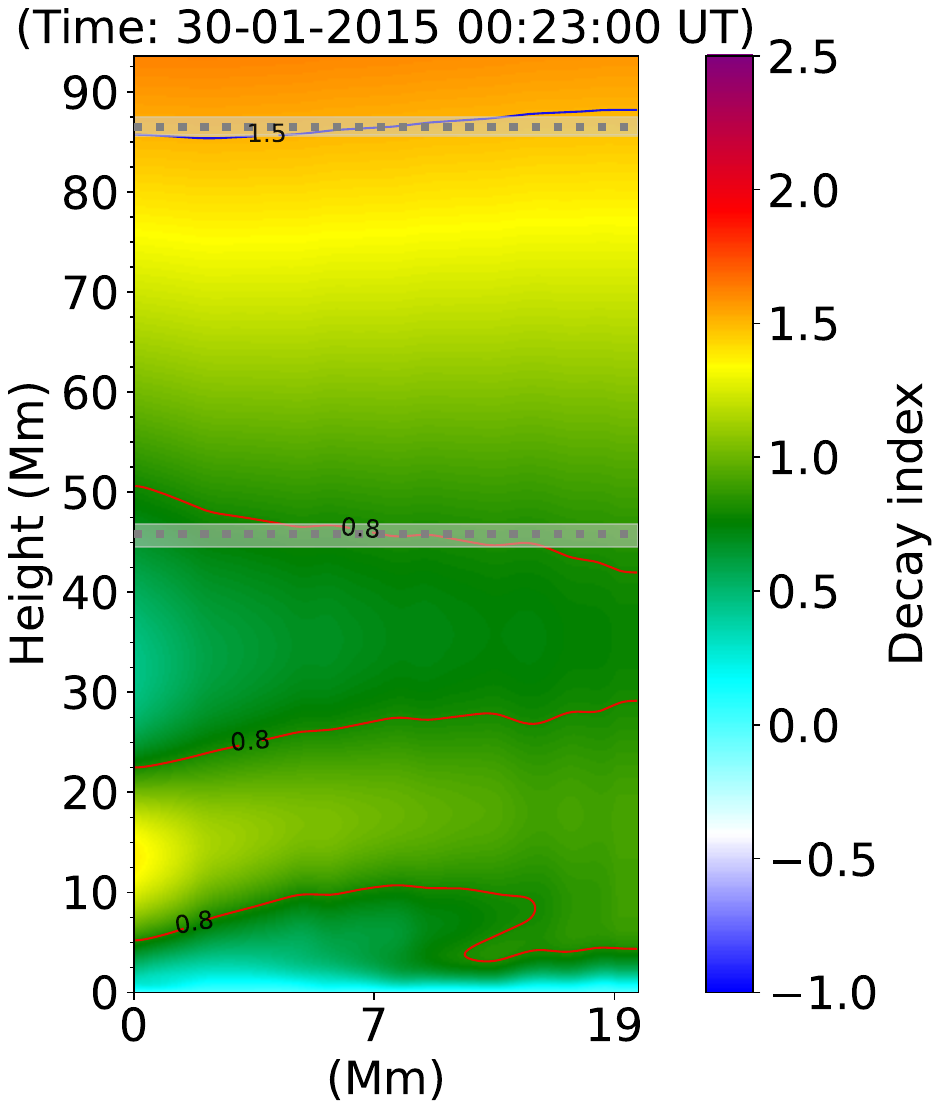}
        \put(-190,107){c) AR 12268}
    \end{subfigure}
    \begin{subfigure}[ht]{0.45\textwidth}
        \centering
        \includegraphics[width=0.8\linewidth,trim={0cm 0cm 0cm 0cm},clip]{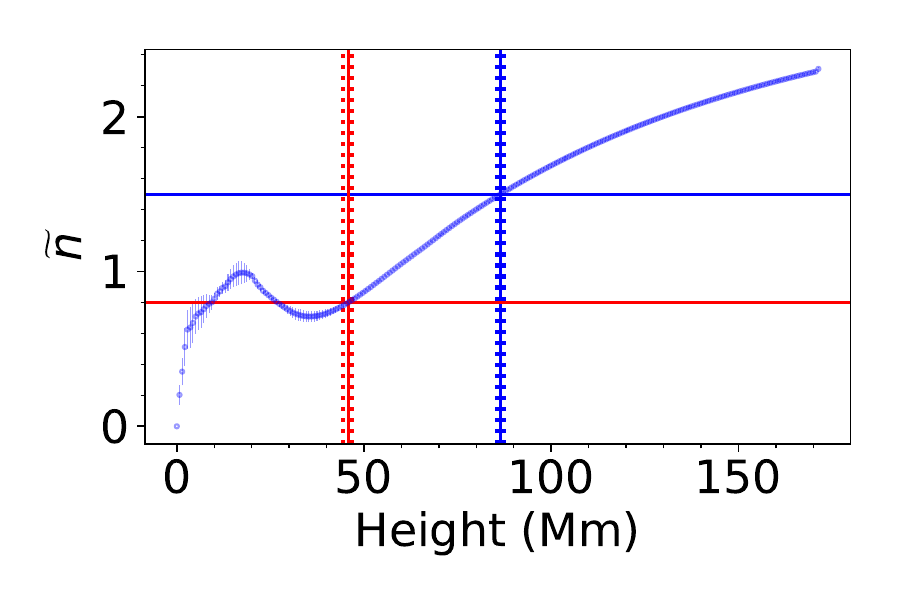}
        \put(-200,100){d)}
    \end{subfigure}
    
    \hspace{0.5cm}\begin{subfigure}[ht]{0.45\textwidth}
        \centering
        \includegraphics[width=0.9\linewidth,trim={0cm 0cm 0cm 1cm},clip]{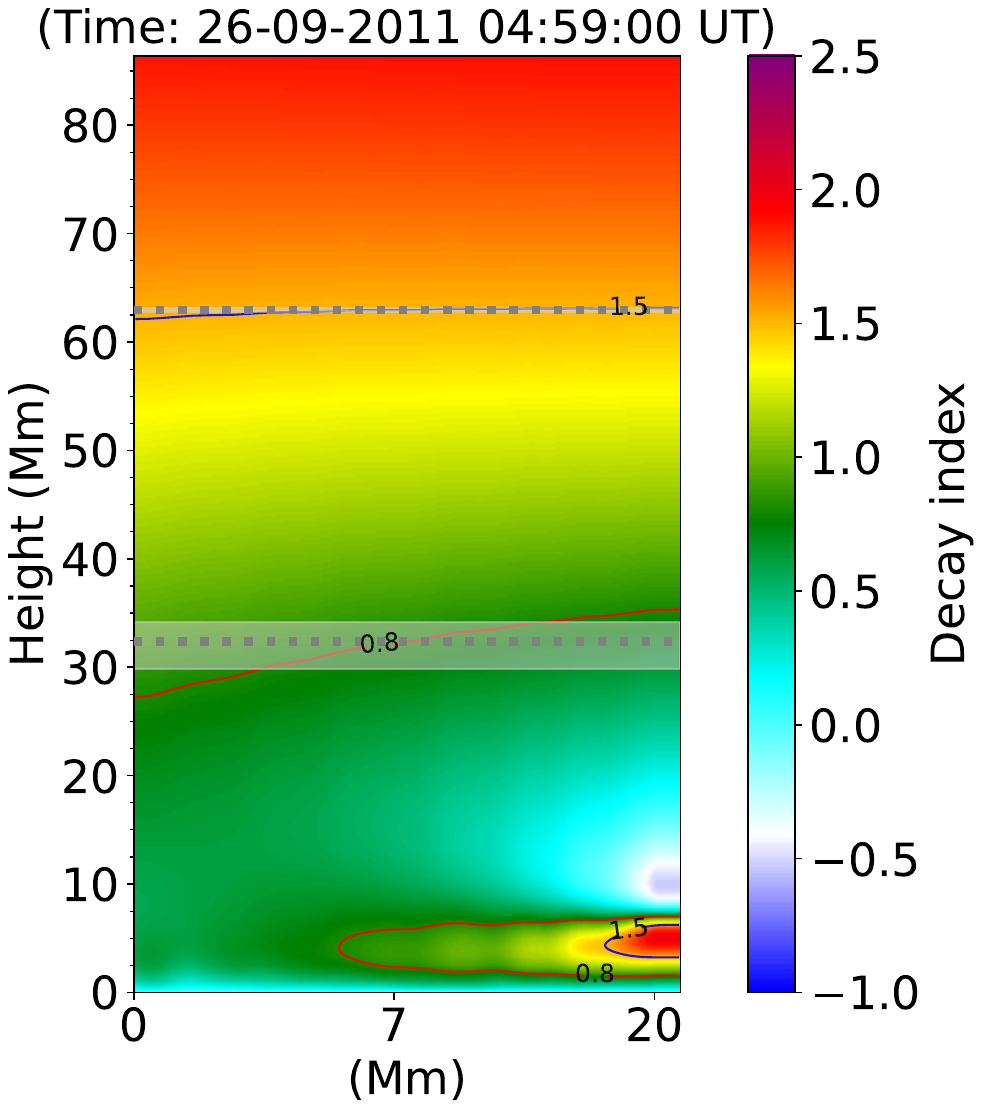}
        \put(-190,107){e) AR 11302}
    \end{subfigure}
    \begin{subfigure}[ht]{0.45\textwidth}
        \centering
        \includegraphics[width=0.8\linewidth,trim={0cm 0cm 0cm 0cm},clip]{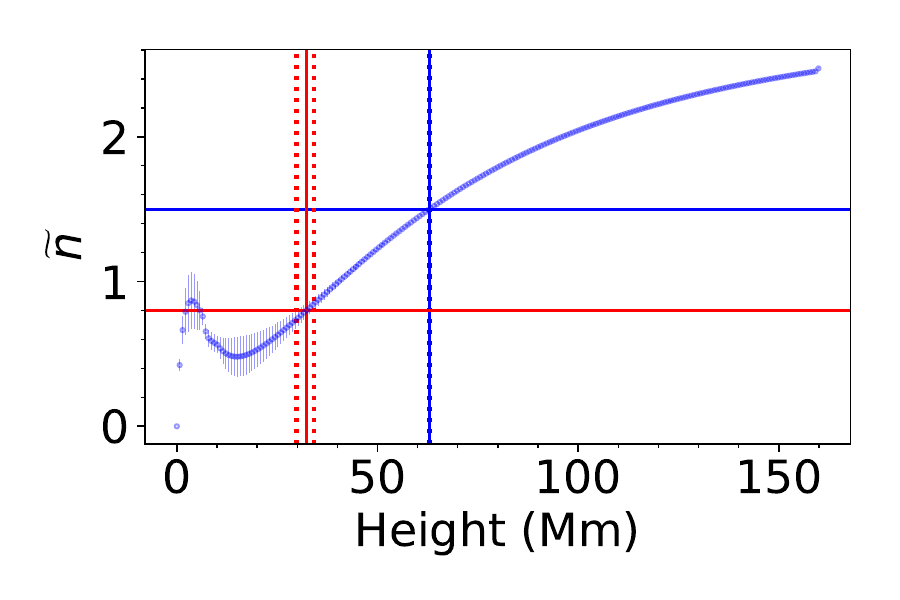}
        \put(-200,100){f)}
    \end{subfigure}
    
    \vspace{-0.2cm}\hspace{0.75cm}\begin{subfigure}[ht]{0.45\textwidth}
        \centering
        \includegraphics[width=0.8\linewidth,trim={0cm 0cm 0cm 2cm},clip]{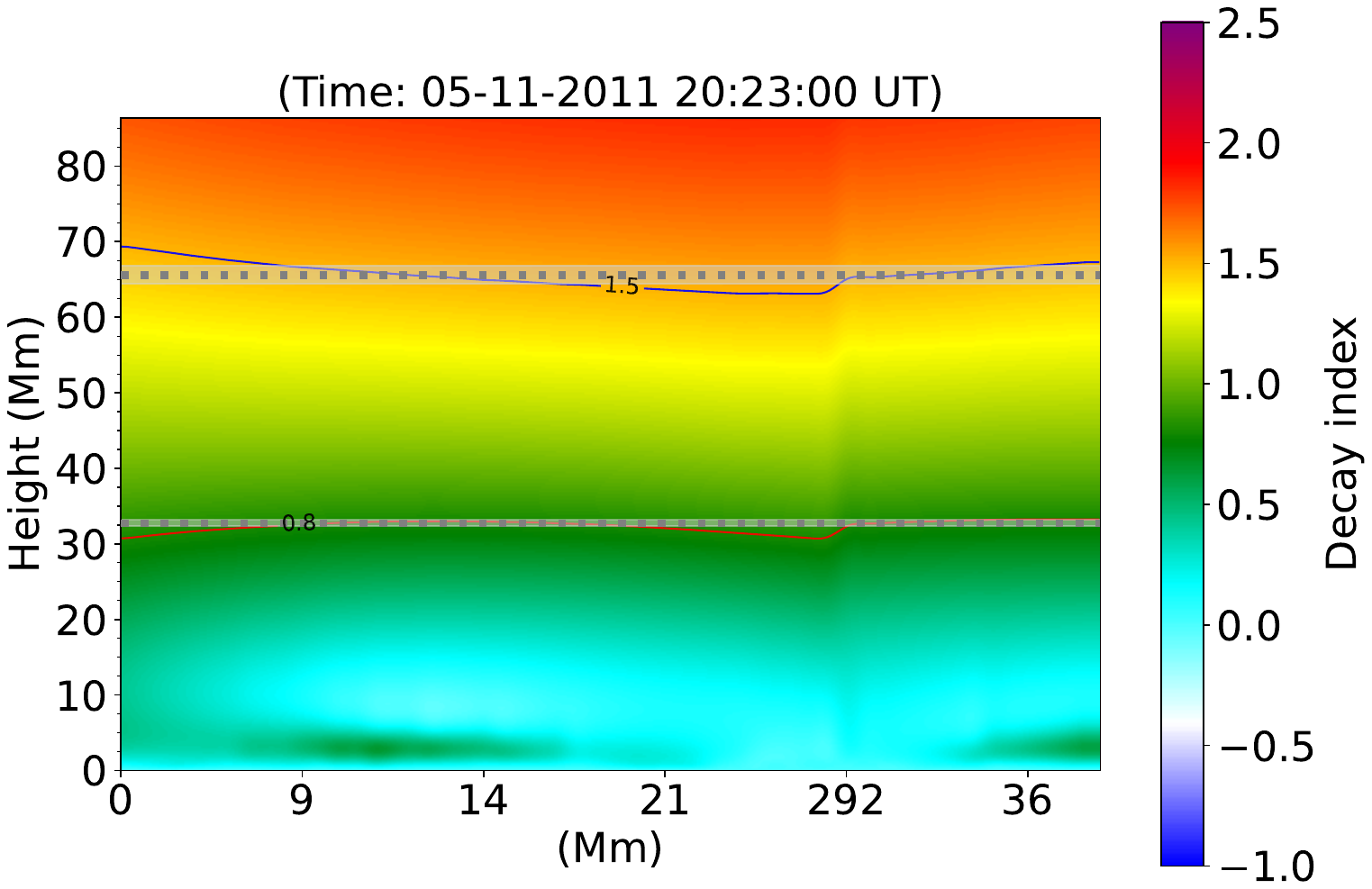}
        \put(-185,115){g) AR 11339}
    \end{subfigure}\hspace{-1em}  
    \begin{subfigure}[ht]{0.45\textwidth}
        \centering
        \includegraphics[width=0.8\linewidth,trim={0cm 0cm 0cm 0cm},clip]{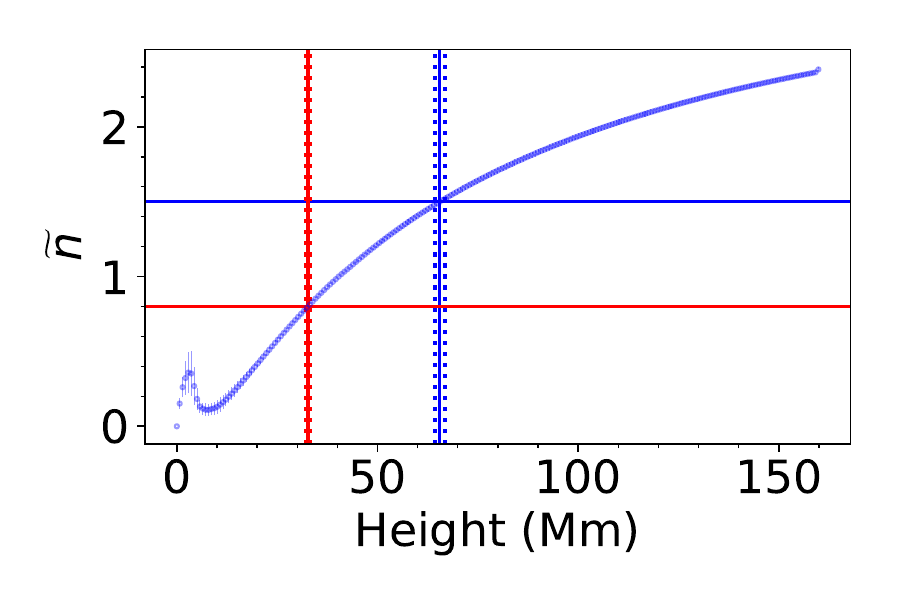}
        \put(-200,100){h)}
    \end{subfigure}
    
    \vspace{-0.2cm}\hspace{0.5cm}\begin{subfigure}[ht]{0.45\textwidth}
        \centering
        \includegraphics[width=0.8\linewidth,trim={0cm 0cm 0cm 1cm},clip]{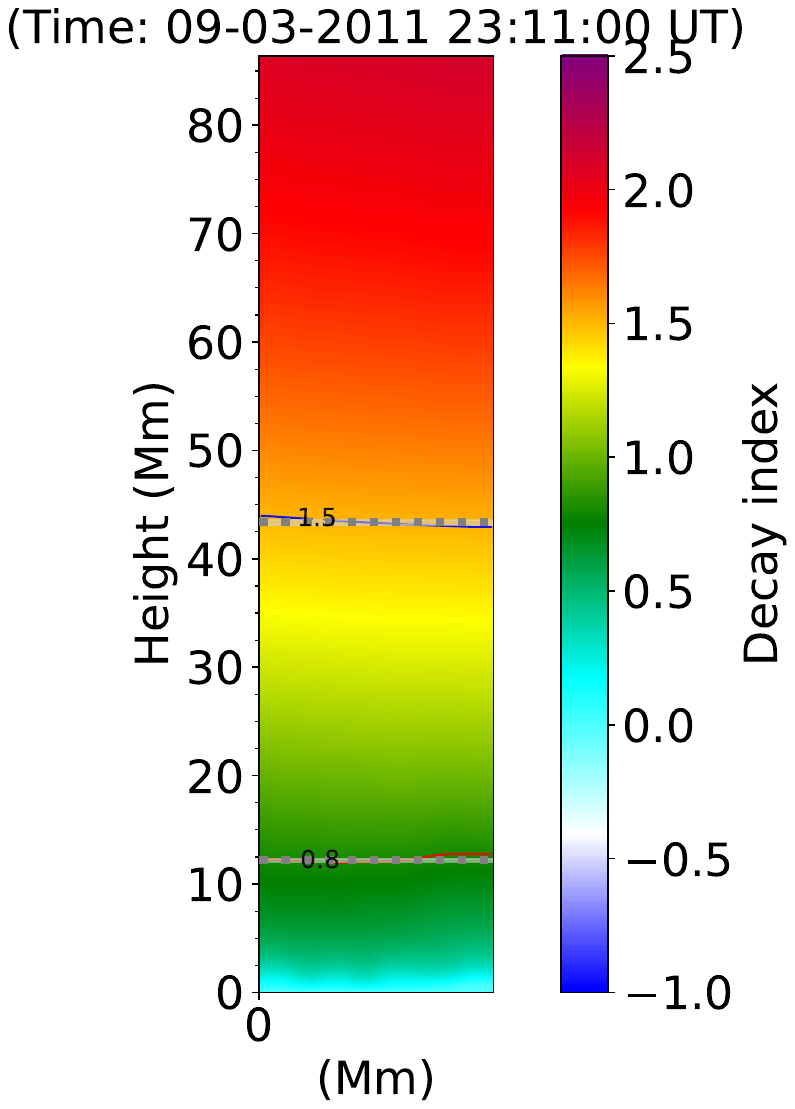}
        \put(-175,107){i) AR 11166}
    \end{subfigure}
    \begin{subfigure}[ht]{0.45\textwidth}
        \centering
        \includegraphics[width=0.8\linewidth,trim={0cm 0cm 0cm 0cm},clip]{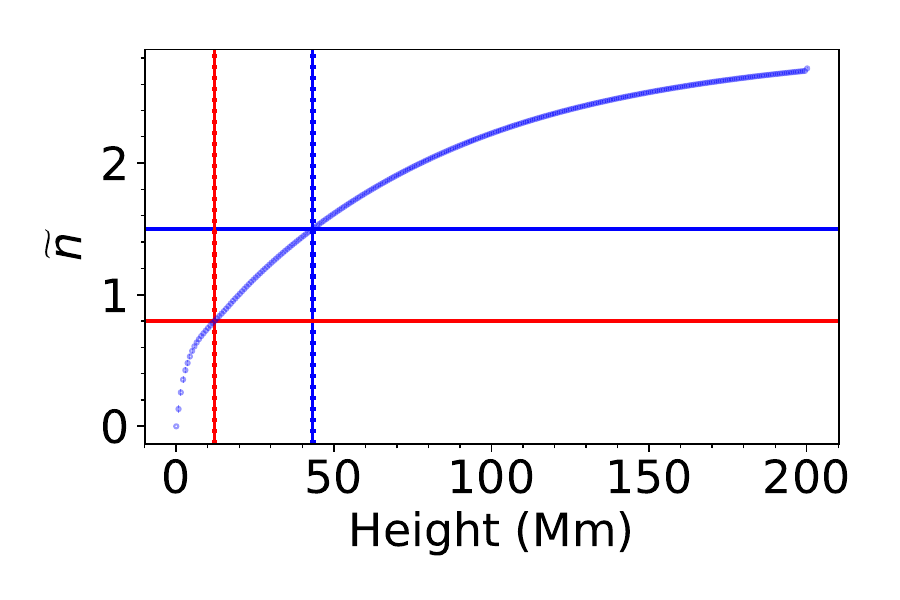}
        \put(-200,100){j)}
    \end{subfigure}
     \caption{
     Distribution of the decay index in the FPIL-aligned planes (left column) and profile of \mb{$\mbf{\nmedian}$} vs.\ height (right column). Layout as in \href{app1_1}{\rrff{Fig.}~\ref{app1_1}.}
     }
    \label{app1_2}
\end{figure*}

\begin{figure*}[htp]
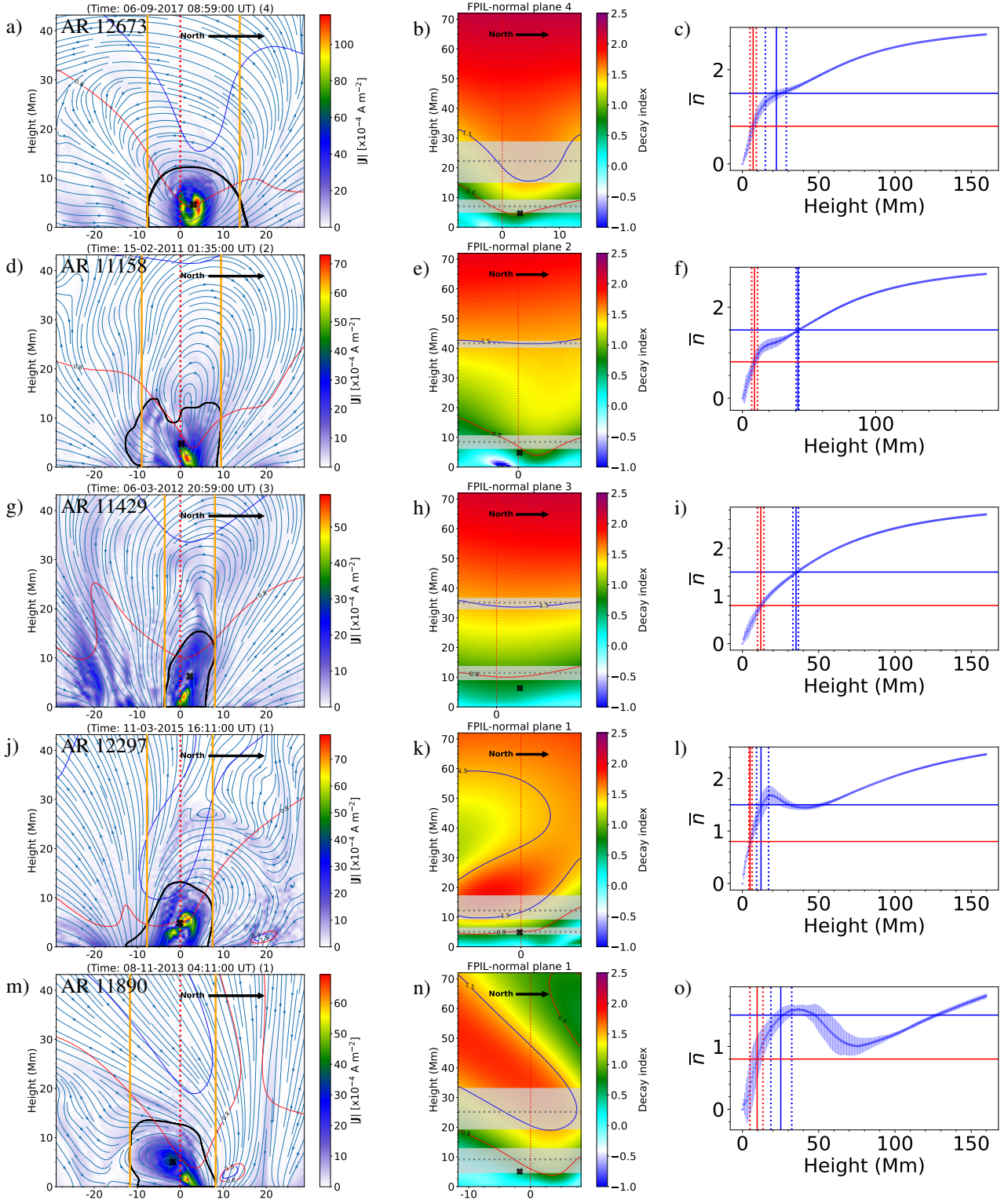

    \captionsetup{width=\linewidth}
    \jmapfiga{norm_width_58.32Mm_Bproj_noise_5G_j_normal_to_pil_}{jmap}{norm_width_58.32Mm_decay_normal_to_pil_}{norm_width_58.32Mm_normal_median_decay_index1d_vs_height}
    \caption{
    \rf{{\it Left column:}} \rf{In-plane distribution of the electric current density (color-coded) and magnetic field (blue stream lines) in selected FPIL-normal planes. Black closed contours mark the regions of strongest electric current \rf{(95th percentile of the current density in plane)}, used to locate the current-weighted center of MFR (SA) \rff{(black cross). The} red vertical dotted line marks the location of the FPIL. Orange vertical lines mark the extent within the FPIL-normal plane within which the decay index is analyzed. Blue (red) contours indicate where $n$ exceeds 1.5 (0.8).}
    \textit{Middle column:} 
    Decay index on selected FPIL-normal planes \rf{covering the FOV indicated by vertical orange lines in the respective panels of the left column. The gray dotted line and shaded region indicates $\mbf{\hcbar}$ (using $\nc$\,=\,1.5 and \mb{$\nc$\,=\,0.8}) and the corresponding spread when averaged over all considered FPIL-normal planes. Layout as in left column of \href{app1_1}{Fig.~\ref{app1_1}}.} 
    {\textit{Right column:}} 
    $\mbf{\nbar}$ vs.\ height profiles derived from all of the FPIL-normal planes \rf{(their footprints are indicated in the respective panels of \href{app1}{\rrff{Fig.}~\ref{app1}}). Layout as in right column of \href{app1_1}{Fig.~\ref{app1_1}}.}
    }
    \label{jmapa}
\end{figure*}

\begin{figure*}[htp]
    \captionsetup{width=\linewidth}
    \vspace{0.4cm}
    \jmapfigb{norm_width_58.32Mm_Bproj_noise_5G_j_normal_to_pil_}{jmap}{norm_width_58.32Mm_decay_normal_to_pil_}{norm_width_58.32Mm_normal_median_decay_index1d_vs_height}
    \caption{Layout as in \href{jmapa}{\rrff{Fig.}~\ref{jmapa}}.}
    \label{jmapb}
\end{figure*}

\end{appendix}

\end{document}